\documentclass[prb,amsfonts,amssymb,floats,twocolumn,superscriptaddress,aps]{revtex4-2}
\usepackage[utf8]{inputenc}
\usepackage[T1]{fontenc}
\usepackage{amsmath}
\usepackage{float}
\usepackage[caption = false]{subfig}
\usepackage{color}
\usepackage{braket}
\usepackage{bm}
\usepackage{dsfont}
\usepackage{adjustbox}
\usepackage{graphicx}
\usepackage{enumitem}
\usepackage{bbm}
\usepackage{tikz}
\usetikzlibrary{decorations.markings}

\usepackage{ytableau}
\ytableausetup{smalltableaux}

\usepackage[colorlinks=true]{hyperref}
\hypersetup{linkcolor=blue,citecolor=blue,urlcolor=blue}

\definecolor{burntorange}{rgb}{0.8, 0.33, 0.0}

\newcommand{\zx}[1]{#1}
\newcommand{\ufps}[1]{#1}

\renewcommand{\Re}{\operatorname{Re}}
\renewcommand{\Im}{\operatorname{Im}}

\newcommand{\iu}{\mathrm{i}} 
\newcommand{\eu}{\mathrm{e}} 
\newcommand{\du}{\mathrm{d}} 
\newcommand{\hc}{\mathrm{h.c.}} 

\newcommand{\bvec}[1]{{\bm{#1}}}

\newcommand{\js}{J_\mathrm{s}}
\newcommand{\jv}{J_\mathrm{v}}

\newcommand{\fs}{f_\mathrm{s}}
\newcommand{\fv}{f_\mathrm{v}}

\newcommand{\Ztwo}{\mathbb{Z}_2}
\newcommand{\Uone}{\mathrm{U(1)}}
\newcommand{\SUtwo}{\mathrm{SU(2)}}
\newcommand{\SUfour}{\mathrm{SU(4)}}

\newcommand{\SOsix}{\mathrm{SO(6)}}

\newcommand{\SO}{\mathrm{SO}}
\newcommand{\SU}{\mathrm{SU}}

\newcommand{\qed}[1]{QED$_3$}

\newcommand{\be}{\begin{equation}}
\newcommand{\ee}{\end{equation}}
\newcommand{\x}{\vec{x}}
\renewcommand{\u}{\vec{u}}
\newcommand{\R}{\vec{R}}

\newcommand{\K}{\vec{K}}
\newcommand{\q}{\vec{q}}
\renewcommand{\a}{\vec{a}}
\newcommand{\Q}{\vec{Q}}
\renewcommand{\k}{\vec{k}}

\newcommand{\mf}{_\mathrm{mf}}
\newcommand{\bk}[1]{\langle #1 \rangle}
\newcommand{\0}{_{\mathrm{QED}_3}}

\graphicspath{{./}{figs/}}

\begin{document}
\tikzset{->-/.style={decoration={
  markings,
  mark=at position #1 with {\arrow{>}}},postaction={decorate}}}
  
\title{Twisted bilayer U(1) Dirac spin liquids}

\author{Zhu-Xi Luo}
\thanks{These two authors contributed equally.}
\author{Urban F. P. Seifert}
\thanks{These two authors contributed equally.}
\affiliation{Kavli Institute for Theoretical Physics, University of California, Santa Barbara, CA 93106}

\author{Leon Balents}
\affiliation{Kavli Institute for Theoretical Physics, University of California, Santa Barbara, CA 93106}
\affiliation{Canadian Institute for Advanced Research, Toronto, ON, Canada M5G 1M1}

\date{\today}

\begin{abstract}
When two layers of two-dimensional materials are assembled with a relative twist, moiré patterns arise, inducing a tremendous wealth of exotic phenomena. In this work, we consider twisting two triangular lattices hosting Dirac quantum spin liquids. \ufps{A single decoupled layer is described by compact quantum electrodynamics in 2+1 dimensions (QED$_3$) with an emergent $\Uone$ gauge field, which is assumed to flow to a strongly interacting fixed point in the IR with conformal symmetry.} 
We use \zx{recent results for the quantum numbers of monopole operators, which tunnel $2 \pi$ fluxes of the compact gauge field.}
\ufps{It is found that, in the bilayer system, interlayer monopole tunneling is a symmetry-allowed relevant perturbation which induces an (ordering) instability. 
We show using perturbation theory that upon twisting the two layers the system remains unstable under the interlayer interaction, but any finite twist angle softens this instability compared to the untwisted case.
To analyse the resulting phase induced by the (twisted) interlayer tunneling, we use ``conformal mean field theory'', which reduces the interacting bilayer system to two copies of QED$_3$ coupled to background fields which are to be determined self-consistently.
In the weak-coupling regime, where the interlayer coupling is weak compared to the energy scale set by the moiré lattice constant, we solve the self-consistency equations perturbatively.
In the limit of strongly coupled layers, a local scaling approximation is used, and we find that the magnetically ordered state exhibits a lattice of magnetic vortices, with the lattice constant tunable through the twisting angle.}
\end{abstract}

\maketitle


\section{Introduction}

\subsection{Motivation}

Van der Waals materials are two-dimensional atomic crystals with strong in-plane covalent bonds and weak interlayer van der Waals (vdW) interactions, which can be exfoliated down to monolayer.
In recent years, numerous experiments have strongly advanced capabilities to prepare and control these materials with atomic precision, and assemble them like quantum Legos \cite{geim2013van,doi:10.1126/science.aac9439}. 

A particularly fruitful avenue in this regard consists in exploiting moiré physics, which arises from relative twisting or lattice mismatches between different layers, leading to quantum interferences \ufps{which quench} the energy scales in the system, often permitting a strongly-coupled, interaction-dominated regime.
One leading example of this manipulation of electronic properties, or twistronics  \cite{PhysRevB.95.075420} in vdW heterostructures, is that of the twisted bilayer graphene \cite{cao2018correlated,cao2018unconventional}, where exotic superconductivity and correlated insulating behaviors have been observed. 

Recently, the study of moiré heterostructures has been extended to the systems with magnetic ordering, where interesting non-collinear magnetic ordering, magnon behaviors and topological spin textures have been found \cite{Hejazi10721,PhysRevLett.125.247201, xu2021coexisting, doi:10.1126/science.abj7478, PhysRevB.102.094404, skyrmion, PhysRevB.103.L140406, 2021skm, PhysRevResearch.3.013027,  PhysRevB.104.L100406}.
A natural follow-up question is then: what happens when one twists quantum spin liquids, i.e. strongly correlated spin systems that intrinsically lack magnetic ordering \cite{Savary_2016}? 

In this work, we will focus on the twisted bilayer of $\Uone$ Dirac spin liquids (DSL) described in the low-energy limit by $N=4$ flavors of Dirac fermions coupled minimally to an emergent $\Uone$ gauge field.
This effective theory is known as QED$_3$, quantum electrodynamics in 2+1 dimensions.
In the context of $\Uone$ DSL, the Dirac fermions are fractionalized quasiparticles which carry the spin degrees of freedom of the electrons, thereby named spinons.
In the absence of monopole and $\SU(N)$ symmetry breaking (which is avoided for a sufficiently large number of fermion flavors $N$), QED$_3$ is assumed to be stable and to flow to a strongly coupled conformal fixed point in the IR \cite{karthik16a,karthik16,kogut04,di2017scaling,Zhijin1,Zhijin2}.

Our motivation for focusing on $\Uone$ DSL is two-fold:
On the one hand, there is evidence which suggests that such a state could be the ground state in realistic and experimentally relevant microscopic spin models, thus constituting a prime example of an exotic highly-entangled strongly-correlated magnetic state of matter.
Dirac spin liquids were originally studied in the context of high-Tc superconductors \cite{RevModPhys.78.17} on the square lattice, and later examined on other lattices  \cite{PhysRevB.63.014413,hermele,PhysRevLett.98.117205,PhysRevB.77.224413} as well.
In particular, on the triangular lattice with both nearest and next nearest neighbor spin couplings, there is considerable numerical evidence suggesting the presence of a Dirac spin liquid  \cite{PhysRevB.92.041105,PhysRevB.93.144411,PhysRevB.92.140403, PhysRevB.95.035141}.
Material candidates have been proposed as well, including Ba$_8$CoNb$_6$O$_{24}$ \cite{PhysRevMaterials.2.044403} and $1$T-TaS$_2$ \cite{Law6996}, the latter being a van-der-Waals material.

On the other hand, $\Uone$ DSLs have commonly been described as ``parent states'' of competing orders, as they offer a unified framework to describe various (seemingly unrelated) magnetic (and valence-bond paramagnetic) ordered states \cite{hermele} by inducing instabilities of the $\Uone$ DSL.
Our present study hence can also be understood as exploring the viability of the concept of the DSL as a parent state, both in homogenous layered systems and upon spatial modulations of the interlayer coupling as an additional control knob, potentially stabilizing various exotic competing ordered states with moiré supermodulations.  


For the spin liquid on the triangular lattice, recent works by Song et al. have shown that all monopoles in the effective QED$_3$ theory carry nontrivial symmetry quantum numbers \cite{song19,song20}.
The lowest-order symmetry-allowed term is a triple monopole \cite{song19,song20}, very likely to be irrelevant based on large-$N$ expansion \cite{dyer13}.
There can also be four fermion terms that break the $\SU(4)$ flavor symmetry, which are found to be irrelevant in $\epsilon$-expansion \cite{di2017scaling}, but relevant in large-$N$ expansions \cite{PhysRevB.97.195115}.
While the stability of the $\Uone$ Dirac spin liquid on the triangular lattice is hence an open question, any instabilities inherent in a single two dimensional layer are relatively weak, if they are present at all.  By contrast, we observe in this work a strong instability (associated with highly relevant inter-layer interactions) of the bilayer system, which can be tuned through twisting.   



\begin{figure}
    \centering
    \includegraphics[width=\columnwidth]{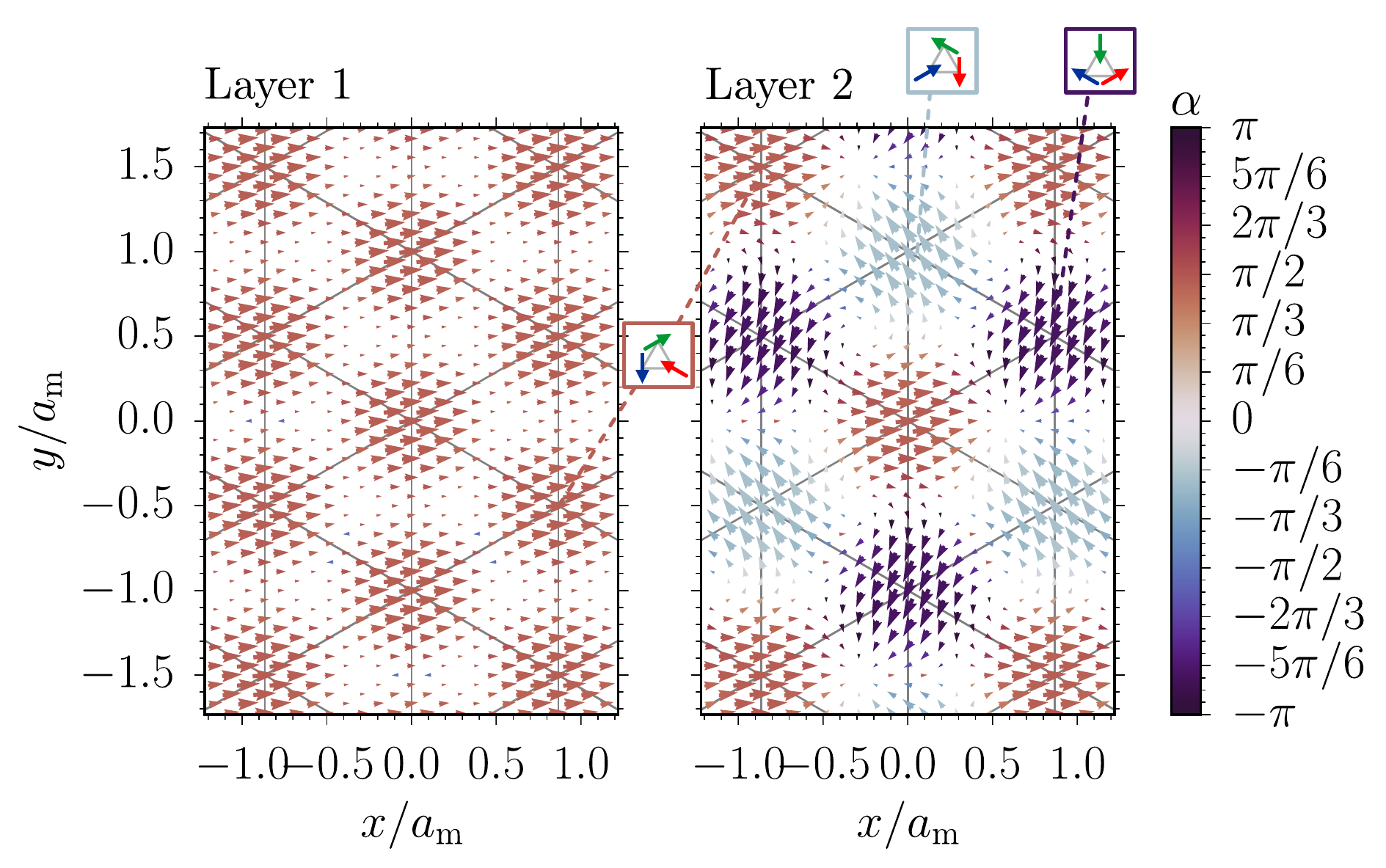}
    \caption{Schematic example order parameters $\langle \bvec{\Phi}_l \rangle$ (on the moiré lattice, indicated by grey lines) for the two layers in the strong coupling limit, where $J_\mathrm{s} \gg a_\mathrm{m}^{2 \Delta_\Phi-3}$. The length of the plotted arrows correspond to the magnitude $|\langle \bvec{\Phi}_l \rangle|$, the direction of the plotted arrows corresponds to $\Re[\langle \bvec{\Phi}_l \rangle/ |\langle \bvec{\Phi}_l \rangle|]$, and the angle $\alpha \equiv \arg \, \Im[\langle \bvec{\Phi}_l \rangle/ |\langle \bvec{\Phi}_l \rangle|]$ is color-coded. The insets depict the \emph{spin configurations} on the A (blue), B (red) and C (green) sublattices of the parent layers, determined via Eq.~\eqref{eq:spin_spin}. As visible, the magnetic order parameter in both layers vanishes at the centers of the moiré triangles, leading to \emph{moiré vortices} with a non-trivial winding number of the relative angle of the order parameters between the two layers. Note that this configuration spontaneously breaks the interlayer exchange symmetry.}
    \label{fig:vortex_lattice}
\end{figure}

\subsection{Summary of results and outline}
\label{sec:preview}

We briefly summarize the model and our results in this subsection.
In each layer, the effective low energy theory is that of the $N_f=4$ \qed{}, i.e., four flavors of Dirac fermions coupled to a $\Uone$ gauge field, which we assume to flow to a conformal fixed point \cite{karthik16a,karthik16,kogut04,di2017scaling}, facilitating our use of conformal data to constrain correlation functions and construct effective actions.

We consider the interlayer coupling of fermion bilinear masses in the two layers as well as interlayer tunneling of monopoles of the $\Uone$ gauge fields. 
Guided by latest conformal bootstrap results \cite{he21,alba22}, we take the monopole tunneling terms to be the most relevant.
For a relative rigid twist of angle $\theta$, symmetry analysis predicts monopole tunneling terms to be of the form 
\be
\begin{split}
 \mathcal{L}_1= & \jv \sum_{a=1}^3 \fv^a(x) \Phi_{1a}^\dagger \Phi_{2a}+\js \fs(x)   \sum_{a=4}^6 \Phi_{1a}^\dagger \Phi_{2a}+\hc,\\
& \fv^a(x)=\eu^{\iu \q_a\cdot \x/2},\quad \text{and} \quad \fs (x)=\sum_{i=1}^3 \eu^{\iu \q_i\cdot \x},
\end{split}
\ee
where $\Phi_{la}$ is the monopole annihilation operator in layer $l$ with flavor $a$.
There are altogether six different flavors, which fall into two classes labeled by the
indices $\mathrm{v}$ ($a=1,2,3$) and $\mathrm{s}$ ($a=4,5,6$) standing for valence bond solid (VBS) and spin channels, respectively.
Proliferating the corresponding monopoles (or a linear combination thereof) yields a corresponding VBS or magnetic order parameter.
The finite twist angle leads to supermodulations of the interlayer couplings, i.e. the functions $f_\alpha(x)$ are periodic on the \emph{moiré scale} with moiré reciprocal vectors $\q_a=\theta \hat{{z}}\times \K_a$, where $\K_a$ is the three $C_3$-symmetry-related Brillouin zone corner vectors of the parent layers. 

{We first analyze the interlayer interaction using perturbation theory, using the fact that monopole correlation functions are strongly constrained by conformal symmetry.   For $\theta=0$, one can predict simply from the scaling dimension $\Delta_\Phi$ that this perturbation theory is divergent, and the uniform coupling of layers is a relevant perturbation in the renormalization group sense.  This divergence appears already at quadratic order in the couplings. For $\theta>0$, the couplings oscillate spatially and have zero mean.  This immediately eliminates any divergence arising at quadratic order, which suggests that, keeping $\theta$ fixed and taking $J_v, J_s$ arbitrarily small, the coupling of layers might be irrelevant.  

A key result of our analysis is that this is not the case.  We find that even for $\theta>0$, a perturbative divergence persists, but appears only at higher order in the couplings.    This leads to a breakdown of perturbation theory which we identify as an instability.  That is, even at non-zero $\theta$ (finite $a_m$), we find that the twisted bilayer system is unstable in the thermodynamic limit (i.e. sample size) $L\to\infty$ for any infinitesimally small $J$.  The transfer of the instability from quadratic to higher order with the introduction of a twist, leads to the conclusion that the instability is \emph{softened} upon introducing a non-zero twist angle $\theta$ -- in this case, the power-law finite-size scaling of the critical $J_c$ reads
\begin{equation} \label{eq:critical-scaling-intro}
    J_c \sim (a_\mathrm{m} L)^\ufps{-(3/2-\Delta_\Phi)},
\end{equation}
compared to $J_c \sim L^{-(3-2 \Delta_\Phi)}$ in the homogeneous case (absence of any twist angle).
Hence, the spatial modulation of the interlayer coupling due to the twisting effectively \emph{renders the interlayer coupling less relevant}.

To study the nature of the phase resulting from the instability,} we make a mean-field approximation by replacing the interlayer interactions with a mean-field action of the form
\begin{equation}
    \Phi_{1a}^\dagger \Phi_{2a} + \hc \to \langle \Phi_{1a}^\dagger \rangle_\mathrm{mf} \Phi_{2a} +  \Phi_{1a}^\dagger \langle \Phi_{2a} \rangle_\mathrm{mf} + \hc,
\end{equation}
where the mean-fields $\langle \Phi_{l,a}\rangle$ are to be determined self-consistently.  This approach is in direct analogy to ``chain mean field theory'' \cite{PhysRevLett.77.2790,PhysRevLett.93.127202,PhysRevB.82.014421} used very successfully to describe coupled one-dimensional spin chains.

Crucially, the presence of both an energy and length scale, namely the interlayer coupling $J$ and the moiré lattice scale, allows us to distinguish between \emph{weak-coupling} $J \ll a_\mathrm{m}^{2 \Delta_\Phi-3}$ and \emph{strong-coupling} $J \gg a_\mathrm{m}^{2 \Delta_\Phi -3}$ regimes.
Here, $\Delta_\Phi \simeq 1.02$ is the scaling dimension of the monopole operator at the \qed{} fixed point, estimated from large-$N$ and bootstrap methods \cite{alba22}.

\zx{On one hand, considering the mean-field theory in the} weak-coupling limit, one may use conformal perturbation theory to quadratic order to obtain above monopole expectation values in each layer, and perform a qualitative analysis of quartic terms to fix accidental degeneracies. {We find that in this perturbative regime, non-trivial solutions to the mean-field equations emerge when $J\sim (a_\mathrm{m} L)^{-\left(\frac{3}{2} - \Delta_\Phi \right)}$, which is precisely the same scaling as obtained in perturbation theory without any mean-field approximation, see Eq.~\eqref{eq:critical-scaling-intro}.}

{To determine the form of the order parameter in the ordered phase, we first consider spin monopoles $\{\Phi_{l4}, \Phi_{l5}, \Phi_{l6}\}$, which are generated by spin-spin interactions in a microscopic theory and are thus expected to be dominant.} At the finite-size critical point, the spatial dependence of the monopole expectation values (which correspond to the Néel order parameters on the two parent triangular lattices) read 
\be
\begin{split}
& \bk{\bm{\Phi}_{1}(\x)}\mf \approx -\frac{A}{|\js|}\tilde{\bm{h}}(0) -\frac{\fs(x)}{3\js }\tilde{\bm{h}}(0),\\
& \bk{\bm{\Phi}_{2}(\x)}\mf \approx  \frac{A}{\js}\tilde{\bm{h}}(0)+\frac{\fs^*(x)}{ 3|\js|}\tilde{\bm{h}}(0).\\
\end{split}
\label{eq:summary_weak}
\ee
Here $\tilde{\bm{h}}(0)$ is a constant three-dimensional complex vector.
$A \sim L/a_\mathrm{m} \gg 1$ is a constant proportional to the ratio between the IR cutoff (sample size $L$) and the moiré lattice constant ($a_\mathrm{m} \sim a_0/\theta$). Therefore, the first uniform terms in the two equations above dominate, and there are some corrections with modulations of moiré scale.
The coupling of VBS monopoles $\{\Phi_{l1},\Phi_{l2},\Phi_{l3}\}$ can be analyzed similarly. 

On the other hand, in the strong-coupling regime of $J_a \gg a_\mathrm{m}^{2 \Delta_{\Phi}-3}$, the spatial modulations of the tunneling amplitudes are slow enough such that locally, the bilayer system can be well-approximated by that of a uniform stacking.
This limit therefore facilitates a scaling ansatz for the mean-field free energy from which correlation functions can be determined (the resulting local degeneracy is lifted by symmetry-allowed gradient terms).
Again focussing on the interlayer coupling of spin monopoles, we find that the monopole expectation values -- and thus the magnetic order parameter -- vanish at the centers of the triangular plaquettes in the moiré triangular lattice, precisely where the interlayer-coupling function $\fs(x)$ possesses zeroes, with
$|\langle \bvec{\Phi}_l(x)| \sim |\fs(x)|^{\frac{\Delta_\Phi}{3-2 \Delta_\Phi}}$. 
Around these zeroes, $\fs(x)$ possesses a non-zero winding number (by $\pm 2 \pi$ going anti-clockwise around right/leftward-facing triangles), which leads to the expectation values of the spin monopoles in the two layers to differ by a spatially modulating phase $\bk{\bm{\Phi}_2}\mf/\bk{\bm{\Phi}_1}\mf=\eu^{\iu \arg \fs(x)}$.
Note that this implies that the interlayer exchange symmetry of the system is broken.

We have thus found that, in the limit of strong coupling (to which the \emph{relevant} interlayer interaction inevitably flows), the model \emph{realizes an exotic magnetic vortex lattice on the moiré scale}.
The latter is experimentally tunable by varying the twist angle $\theta$.
An exemplary configuration for such a moiré vortex lattice, showing the order parameter configurations in both layers along with microscopic spin configurations, is depicted in Fig.~\ref{fig:vortex_lattice}.

\zx{One may argue that one could obtain bilayer ordered states qualitatively resembling our results by coupling together order parameters for \emph{classically ordered} phases, such as the $120^\circ$ N\'eel antiferromagnetic phases, in the two layers (resulting bilayer ordered states are expected to be similar to what we found but with explicitly fixed order parameter magnitudes, while within our framework they vary with critical exponents related to operator scaling dimensions of the (conformal) fixed point theory).
However, we emphasize that it is \emph{a priori} unclear if the (twisted) bilayer coupling induces an ordering instability of the two Dirac spin liquids (which, in a single layer, are stable phases).
The problem of stability of the bilayer DSL thus necessitates studying perturbations to the IR fixed point, which is described by a conformal field theory specified by scaling dimensions and an operator algebra. In contrast, a non-linear sigma model formalism for the intertwined order parameters of valence bond solid and magnetic phases would not be able to describe the physics at the fixed point.}

The remainder of this paper is organized as follows.
We review Dirac spin liquids (i.e. the physics of a single layer) and their conformal low-energy field theory in \ref{sec:review}.

In section \ref{sec:interlayer}, we derive a continuum description for the interlayer interactions based on symmetry principles.
{We use perturbation theory to analyze the stability of the system under the (twisted) interlayer interaction in Sec.~\ref{sec:pert-theory}.}
In section \ref{sec:mf}, we develop a mean-field treatment for the interlayer monopole tunneling term, {and solve the resulting mean-field equations perturbatively in Sec.~\ref{sec:perturbativeMFT}, followed by a} strong-coupling analysis in Sec.~\ref{sec:strong}.
Finally, in Sec.~\ref{sec:discussion}, we present a conformal renormalization group analysis for homogeneous interlayer coupling along with the summary and outlook.

\section{$\Uone$ Dirac spin liquids}
\label{sec:review}

In this work, we will be interested in spinons dispersing on the triangular lattice with a staggered $\pi/0$ flux background, leading to two Dirac cones per spin at momenta $\vec{K}_1 = 2 \pi (1/3,-1/\sqrt{3})^\top / a_0$ and $\vec{K}_2 = (-4 \pi/3,0)^\top / a_0$ at zero energy.
In the long-wavelength (continuum) limit, the system can be described in terms of 2+1-dimensional quantum electrodynamics (\qed{}) with the Euclidean Lagrangian
\begin{equation} \label{eq:qed3}
	\mathcal{L}_{\mathrm{QED}_3} = \sum_{i=1}^{N=4} \left[- \bar{\psi}_i \gamma^\mu \left(\partial_\mu - \iu a_\mu \right) \psi_i \right] + \frac{1}{4g^2} f_{\mu \nu} f^{\mu \nu},
\end{equation}
where each $\psi_i$ is a two-component Dirac fermion with $N=4$ flavors (2 spin $\times$ 2 valley degrees of freedom), $\gamma^\mu$ are the gamma matrices, $a_\mu$ denotes the emergent $\Uone$ gauge field with field strength tensor $f_{\mu \nu}$, and $g$ is a coupling constant.
As visible from \eqref{eq:qed3}, the $\SUtwo_\mathrm{spin} \times \SUtwo_\mathrm{valley}$ symmetry becomes enhanced to $\SUfour$ at low energies. 
In contrast to quantum electrodynamics in 3+1 dimensions, the coupling constant $g^2$ has unit mass dimension and thus the theory flows to strong coupling in the IR.
In particular, studies of \qed{} with $N$ flavors have shown that for sufficiently large $N$, the IR fixed point possesses conformal symmetry, while the fate of the theory for small $N$ is uncertain \cite{appel86,bashir08,grover14,janssen14}.

The Dirac fermions in \eqref{eq:qed3} can become gapped upon adding (or spontaneously generating) bilinear masses $M^{\mu \nu} = \bar \psi \sigma^\mu \tau^\mu \psi$, with $\sigma^\mu$ ($\tau^\nu$) denoting Pauli matrices acting on spin (valley) components of the Dirac fermions (note that, as we work in a flat Euclidean spacetime, we are free to use upper/lower indices for notational convenience).
Note that $M^{\mu \nu}$ transforms in a reducible representation of $\SUfour$ which splits as $16 = 1 \oplus 15$ into the singlet and 15-dimensional adjoint irreducible representations.
While adding mass terms explicitly to \eqref{eq:qed3} is forbidden by symmetry, a scenario of spontaneous chiral symmetry breaking has been suggested for a small enough flavor number $N_c$, with the precise nature of the phase structure and the value of $N_c$ under investigation \cite{appel86,bashir08,grover14,janssen14}.
In the following, we will not consider such scenario and instead assume that the theory at $N=4$ flows to the conformal IR fixed point, as supported by recent estimates that give $N_c < 4$ \cite{herbut16,karthik16}.

We further note that in writing the action $S_{\mathrm{QED}_3} = \int \du^3 x \mathcal{L}_{\mathrm{QED}_3}$ as the (continuum) low-energy theory for the bilayer $\Uone$ Dirac spin liquid, one implicitly assumes the presence of a UV cutoff scale given by the (single-layer) lattice spacing $a_0$ beyond which non-universal microscopic (lattice) details of the interacting spin system are of importance.
In real space, the UV cutoff gives a lower bound $a_0\leq |x-y|$ on the separation of two operator insertions at points $x,y$.

\subsection{Monopole operators}
\label{subsec:parent_monopole}

\begin{figure}
    \centering
    \includegraphics[width=.9\columnwidth]{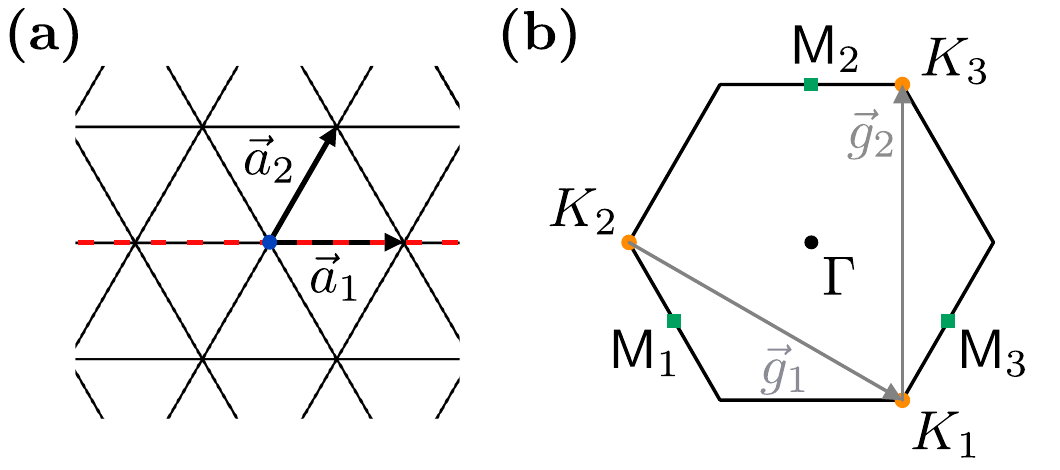}
    \caption{(a) Basis vectors and symmetry operations on the triangular lattice. The blue dot is the rotation center for $C_6$, and the red dashed line is the reflection axis. (b) Hexagonal Brillouin zone on the triangular lattice with the high-symmetry $\Gamma$-,$K$- and $\mathsf{M}$ points and reciprocal lattice vectors $\vec g_1,\vec g_2$.}
    \label{fig:latbz}
\end{figure}

As written, the theory $\mathcal{L}_{\mathrm{QED}_3}$ is endowed with a \emph{topological} conserved current $j^\mu_\mathrm{top} = \epsilon^{\mu \nu \rho} \partial_\nu a_\rho / (2 \pi)$ corresponding to the conservation of the magnetic flux of the emergent $\Uone$ gauge field.
However, this conservation law is an artifact of the continuum limit -- considering that \eqref{eq:qed3} follows from a lattice gauge theory with a \emph{compact} $\Uone$ gauge field, we may consider operators which insert a $2 \pi$ (emergent) magnetic flux or a multiple thereof.
These operators are commonly referred to as monopole operators $\mathcal{M}(x)$, which carry unit charge under the $\Uone_\mathrm{top}$ symmetry.
As these operators cannot be written in terms of the fermions $\psi(x)$ or the gauge-field $a_\mu$, it is convenient to employ the theory's conformal symmetry.
This allows us to characterize these operators through the state-operator correspondence as a scalar primary operator with some scaling dimension $\Delta_\Phi$ which may be evaluated in a controlled manner by quantizing the theory on $S^2 \times \mathbb{R}$ with a given monopole configuration, and performing a large-$N$ expansion \cite{boro02}.
To leading order in $1/N$, this amounts to quantizing the Dirac operator on a sphere which is pierced by a $2 \pi$ magnetic flux, yielding $N=4$ fermionic zero modes (one zero mode per flavor).
By gauge invariance, the four zero modes must be half-filled, so that there are $\binom{4}{2} = 6$ distinct physical states resulting from filling the fermionic zero modes associated with a monopole operator. 
Schematically, these physical monopole operators may be written as \cite{song19,song20}
\begin{equation}
	\Phi_{\alpha \beta} \sim f^\dagger_{\alpha} f^\dagger_{\beta} \mathcal{M}^\dagger,
\end{equation}
where $f_\alpha^\dagger$ are fermionic zero-mode creation operators, so that $\Phi_{\alpha \beta}$ must transform in the 6-dimensional antisymmetric representation of $\SUfour$.
It is convenient to employ the isomorphism $\SOsix = \SUfour / \Ztwo$ such that the monopole operators $\Phi_a$ with $a=1,\dots,6$ are taken to transform in the defining (vector) representation of $\SOsix$.
This also implies that the $\SUtwo_\mathrm{spin},\SUtwo_\mathrm{valley}$ subgroups of $\SUfour$ are isomorphic to the respective $SO(3)_{\text{s/v}}$ subgroups of $\SOsix$. In addition, the monopole operator is odd under both the $\SOsix$ center and a $\pi$-rotation in $U(1)_{\text{top}}$, so the global symmetry group of the low energy theory is $\SOsix \times \Uone_\mathrm{top}/\mathbb{Z}_2$, together with charge conjugation, time reversal $\mathcal{T}$ and Lorentz symmetries. The discrete microscopic (UV) symmetries are shown in Fig.~\ref{fig:latbz} and include translations $T_{1,2}$, reflection $R$, six-fold rotation $C_6$ and time reversal $\mathcal{T}$, which act non-trivially on the operators of the continuum (IR) field theory.
While the transformation properties of the singlet and adjoint masses straightforwardly follow from the microscopic implementation of above symmetries, the appropriate transformations of monopole operators are given by combinations of discrete Lorentz symmetries, $\SOsix$ transformations (due to the zero modes) and $\Uone_\mathrm{top}$ rotations resulting from the microscopic symmetries acting on the filled Dirac sea with a $\pm 2 \pi$ background flux.
In this spirit, the action of microscopic symmetries on the monopole operator has been found and tabulated by Song \emph{et al.} in Refs.~\onlinecite{song19} and \onlinecite{song20} using both numerical methods as well as a study of the band topology of spinons.
For reference, we reproduce their results in Table~\ref{tab:sym}.

We conclude this section by relating the continuum field theory monopole operators to symmetry-equivalent microscopic observables. 
{\renewcommand{\arraystretch}{1.2}
\begin{table}
\caption{\label{tab:sym}Transformation laws of the monopoles under microscopic symmetries. The second/third column represents the transformation under translation along $\vec{a}_1/\vec{a}_2$ in Fig. \ref{fig:latbz}, fourth column the reflection symmetry, fifth column the six-fold rotation symmetry, and last column the time reversal symmetry. This table is adapted from Table 2 in ref. \cite{song19}. }
\begin{ruledtabular}
\begin{tabular}{c|ccccc}
    & $T_1$ & $T_2$ & $R$ & $C_6$ & $\mathcal{T}$ \\ \hline
     $\Phi_1$ & $\eu^{\iu \frac{\pi}{3}} \Phi_1$ & $e^{-i\frac{\pi}{3}} \Phi_1$ & $-\Phi_3$ & $\Phi_2^\dagger$ & $\Phi_1^\dagger$ \\ \hline
     $\Phi_2$ & $e^{-i\frac{2\pi}{3}}\Phi_2$ & $e^{-i\frac{\pi}{3}} \Phi_2$ & $\Phi_2$ & $-\Phi_3^\dagger$ & $\Phi_2^\dagger$ \\ \hline
     $\Phi_3$ & $\eu^{\iu \frac{\pi}{3}}\Phi_3$ & $\eu^{\iu \frac{2\pi}{3}}\Phi_3$ & $-\Phi_1$ & $-\Phi_1^\dagger$ & $\Phi_3^\dagger$ \\ \hline 
     $\Phi_{4/5/6}$ & $e^{-i\frac{2\pi}{3}}\Phi_{4/5/6}$ & $\eu^{\iu \frac{2\pi}{3}}\Phi_{4/5/6}$ & $\Phi_{4/5/6}$ & $-\Phi_{4/5/6}^\dagger$ & $-\Phi_{4/5/6}^\dagger$
\end{tabular}
\end{ruledtabular}
\end{table}}
From Table~\ref{tab:sym}, the first three monopoles $\bk{\Phi_{1,2,3}}$  are time-reversal invariant.
Further imposing spin-rotational invariance, the corresponding microscopic operators at lowest order are of the form $\bm{S}_i\cdot \bm{S}_j,$ consistent with the expectation that these three monopoles describe VBS order. 
The simplest choice that can recover Table~\ref{tab:sym} is
\be
\begin{split}
& \bm{S}_{i}\cdot \bm{S}_{i+\delta_1} = \text{Re}\left[\eu^{\iu \k_1\cdot \vec{R}_i}\ufps{\Phi_1}(\vec{R}_i)\right],\\
& \bm{S}_{i}\cdot \bm{S}_{i+\delta_2} = \text{Re}\left[\eu^{\iu \k_2\cdot \vec{R}_i}\ufps{\Phi_2}(\vec{R}_i)\right],\\
& \bm{S}_{i}\cdot \bm{S}_{i+\delta_3} = -\text{Re}\left[\eu^{\iu \k_3\cdot \vec{R}_i}\ufps{\Phi_3}(\vec{R}_i)\right],
\end{split}
\label{eq:vbs_spin}
\ee
where $\bm{S}_i\cdot \bm{S}_{i+\delta_{j}}$ with $j=1,2,3$ labelling the bonds along the directions $\a_1, \a_2-\a_1$ and $-\a_1-\a_2$, respectively, with
\begin{equation}
	\vec a_1 = a_0 (1,0)^\top \quad \text{and} \quad \vec a_2 = a_0 (1/2,\sqrt{3}/2)^\top
\end{equation}
being two lattice vectors, and the corresponding reciprocal lattice vectors are given by $\vec g_1 = (2\pi,-2 \pi/\sqrt{3})^\top/a_0$ and $\vec g_2 = (4\pi/\sqrt{3},0)^\top/a_0$.

The bonds are separated from each other by $120^\circ$ rotations.
The momenta $\k_i=\ufps{-}\K_i/2$ in \eqref{eq:vbs_spin} are given by half of the Brillouin zone's $K$-points,
\be
\K_{1,3}=\frac{4\pi}{3a_0}\left(\frac{1}{2},\mp\frac{\sqrt{3}}{2}\right)^{\top},\quad  \K_2=\frac{4\pi}{3a_0} (-1,0)^{\top}.
\label{eq:k}
\ee
Here we use the vector symbol $\vec{\cdot}$ to label 2-dimensional vectors in space/reciprocal space, and use boldface to denote $\SO(3)_{\text{v/s}}$ (or $\SO(6)$) vectors.

The vector $\bm{\Phi}_\mathrm{s}=(\Phi_4,\Phi_5,\Phi_6)$ corresponds to the three components the Néel order parameter which determines the spin-density in the magnetically ordered phase as
\be
\bm{S}_{i}=\Re \left[\iu \eu^{\iu \K_1\cdot\R_i}\bm{\Phi}_\mathrm{s}\right].
\label{eq:spin_spin}
\ee
From now on, we will refer to $\{\Phi_{1},\Phi_{2},\Phi_{3}\}$ as ``VBS monopoles'' and $\{\Phi_{4},\Phi_{5},\Phi_{6}\}$ as ``spin monopoles'', as they can be understood as order parameters for valence bond solid and magnetically ordered phases, respectively.

\subsection{Conformal data and operator product expansions} \label{sec:data_OPE}

In general, a conformal field theory is fully specified by its conformal data, containing the operator spectrum as well as the \emph{operator product expansion} (OPE) coefficients. 
The two lowest-lying sets of primaries in the spectrum are given by the 15 adjoint masses $M^{i0}$, $M^{0i}$ and $M_{ij}$ (where $i,j=1,2,3$) as well as monopole operators $\Phi_a$.
Their scaling dimensions to first subleading order in $1/N$ are given by $\Delta_M \approx 1.46$ and $\Delta_\Phi \approx 1.02$ \cite{chester16}.
A recent comprehensive conformal bootstrap study \cite{alba22} finds a striking match (under certain CFT bootstrap assumptions), placing the monopole operator scaling dimension within the range $\Delta_\Phi \in (1.02,1.04)$.
Assuming $\Delta_\Phi = 1.02$ further yields a fermion adjoint mass scaling dimension of $\Delta_M \in (1.33,1.66)$.
Another recent bootstrap work \cite{he21} explicitly accounting for the fate of the UV symmetries of the Dirac spin liquid on the triangular lattice found that stability demands $\Delta_\Phi >1.046$.
Note that Monte Carlo simulations \cite{karthik16} appear to be consistent with the latter study, but have claimed to be ruled out by the aforementioned bootstrap study \cite{alba22}. 

Note that the precise numerical values of $\Delta_\Phi$ and $\Delta_M$ are inconsequential to our study below. However, given the recent results referenced above, it appears to be justified to take $ \Delta_\Phi < \Delta_M \lesssim 1.5$ when analyzing scaling behavior and (ir)relevance of operators.

Crucially, conformal invariance determines the two-point functions of two primary operators $\mathcal{O}_i,\mathcal{O}_j$ up to a global constant, which may be absorbed into the normalization of the operators, 
\begin{equation} \label{eq:twopoint}
    \langle \mathcal{O}_i(x) \mathcal{O}_j(y) \rangle = \frac{\delta_{ij}}{|x-y|^{2 \Delta_{i}}}.
\end{equation}
Further, CFTs admit operator product expansions: two operators approaching each other may be expanded in primaries $\mathcal{O}_k$ as
\begin{equation} 
    \lim_{x \to y} \mathcal{O}_i(x) \mathcal{O}_j(y) = \lim_{x\to y} \sum_k \frac{C_{ij}^k}{|x-y|^{\Delta_i + \Delta_j - \Delta_k}} \mathcal{O}_k(y),
\label{eq:OPE-def}
\end{equation}
where the $C^k_{ij}$ are the OPE coefficients.
Note that, as written, \eqref{eq:OPE-def} corresponds to an asymptotic form of the operator product expansion as $x\to y$, while in general any OPE for distinct $x,y$ can be shown to be a \emph{convergent} series (extending over primaries and descendants) in a conformal field theory, implying that all correlation functions are fully determined by OPE coefficients and scaling dimensions $\Delta_\mathcal{O}$ \cite{pappa12}.

The form of the OPE in \eqref{eq:OPE-def} is strongly constrained by symmetry considerations: If the two operators $\mathcal{O}_i$ and $\mathcal{O}_j$ transform in some irreducible representation of a symmetry group $\mathcal{G}$, the operators on the right-hand side of \eqref{eq:OPE-def} can be classified according to the irreducible representations of $\mathcal{G} \times \mathcal{G}$. 
Using the results by Song \textit{et al.} \cite{song19,song20}, we can hence formulate the operator-product expansions for the most relevant monopole and mass operators up to global constants (with $\sim$ denoting the asymptotic character of the expressions),
\begin{subequations}
\begin{align} \label{eq:phi-phi-OPE}
&\Phi_a^\dagger(x) \Phi_b(y) \sim \frac{\delta_{ab}}{|x-y|^{2 \Delta_\Phi}} + \frac{\iu c_{\Phi\Phi}^M \mathcal{F}^{ab}_{\mu \nu}}{|x-y|^{2 \Delta_\Phi- \Delta_M}} M_{\mu \nu}(y) + \dots  \\
&\Phi_a(x) M_{\mu \nu}(y) \sim \frac{\iu c_{\Phi M}^\Phi \bar{\mathcal{F}}_{ab}^{\mu\nu}}{|x-y|^{\Delta_M}} \Phi_b(y) + \dots, \label{eq:phi-M-OPE}
\end{align}
\end{subequations}
where $\mathcal{F}^{ab}_{\mu \nu}$ is a tensor which maps elements of the 15-dimensional adjoint representation (indexed by $\mu,\nu$, with $\mu = \nu = 0$ excluded from any summation) of $\SUfour$ to the rank-2 antisymmetric representation of $\SOsix$ (indexed by $a,b$), and $\bar{\mathcal{F}}_{ab}^{\mu \nu}$ is the corresponding inverse tensor satisfying $\mathcal{F}_{\mu \nu}^{ab} \bar{\mathcal{F}}_{ab}^{\rho \lambda} = \delta_\mu^\rho \delta_\nu^\lambda$.
An explicit construction of $\mathcal{F}$ shows that one can take $\bar{\mathcal{F}}_{ab}^{\mu \nu} =  \mathcal{F}_{\mu \nu}^{ab}/2$, and we further note the identity $\mathcal{F}^{ab}_{\mu \nu} \mathcal{F}^{cd}_{\mu \nu} = \delta^{ac} \delta^{bd} - \delta^{ad} \delta^{bc}$.
We give details on the derivation of Eqs.~\eqref{eq:phi-phi-OPE} and \eqref{eq:phi-M-OPE} in the Appendix.
The real OPE coefficients $c_{\Phi \Phi}^M$ and $c^{\Phi}_{\Phi M}$ cannot be determined from symmetry considerations, but are constrained by consistency relations encoding the associativity of various four-point functions.
Determining numerical values from these sets of constraints is a key element of the conformal bootstrap program \cite{poryvi19}.

\section{Continuum theory for bilayer $\Uone$ Dirac spin liquids}
\label{sec:interlayer}

In this section, we derive a continuum model for the bilayer Dirac spin liquids with arbitrary elastic deformations.
We start by assuming that the two layers are only slightly deformed with respect to each other, so that the interlayer interaction can be described as the integral of a local Lagrangian density. To make the locality manifest, the following Eulerian coordinates will be used,
\be
\x_l=\R_l+\u_l(\x)+\vec{z}_l,
\ee
where $l\in\{1,2\}$ is the layer index, $\R$ describes the coordinates in the parent triangular lattice, $\u$ is the smooth deformation satisfying ${\partial} \u \ll 1$,  $\x$ is the actual position in the bilayer system, and $\vec{z}$ represents the vertical displacement between the two layers.
In the case of a relative rigid twist of angle $\theta$, we have 
\be
\u_{1/2} =\pm \frac{\theta}{2} \hat{{z}}\times \x.
\ee
As usual \cite{macd14}, such a rigid relative twist leads to the formation of a \emph{moiré superlattice} with reciprocal lattice vectors $\vec g_i^{(\mathrm{m})} = - \theta \hat{z} \times \vec g_i$ , and the corresponding lattice vectors for the moiré superlattice read $\vec a_1^{(\mathrm{m})} = (0,-1)^\top a_0/\theta$ and $\vec a_2^{(\mathrm{m})} = (\sqrt{3}/2,-1/2)^\top a_0/\theta$, which implies that the moiré lattice constant $a_\mathrm{m} = a_0 /\theta$ is inverse proportional to the twist angle $\theta$.

In the following, we will use the physical symmetries to constrain the possible interlayer interactions. 
Considering the scaling dimensions review in Sec.~\ref{sec:review}, two types of interactions are most relevant: (i) interlayer tunneling terms of the monopoles, which keep the \emph{total} emergent magnetic flux conserved in the full bilayer system; and (ii) interlayer couplings of the bilinear masses from the two layers. We will discuss these two cases separately in the following two subsections.

\subsection{Interlayer monopole couplings}
\label{subsec:bootstrap}

\begin{figure}[tbp]
    \centering
    \includegraphics[width=\columnwidth]{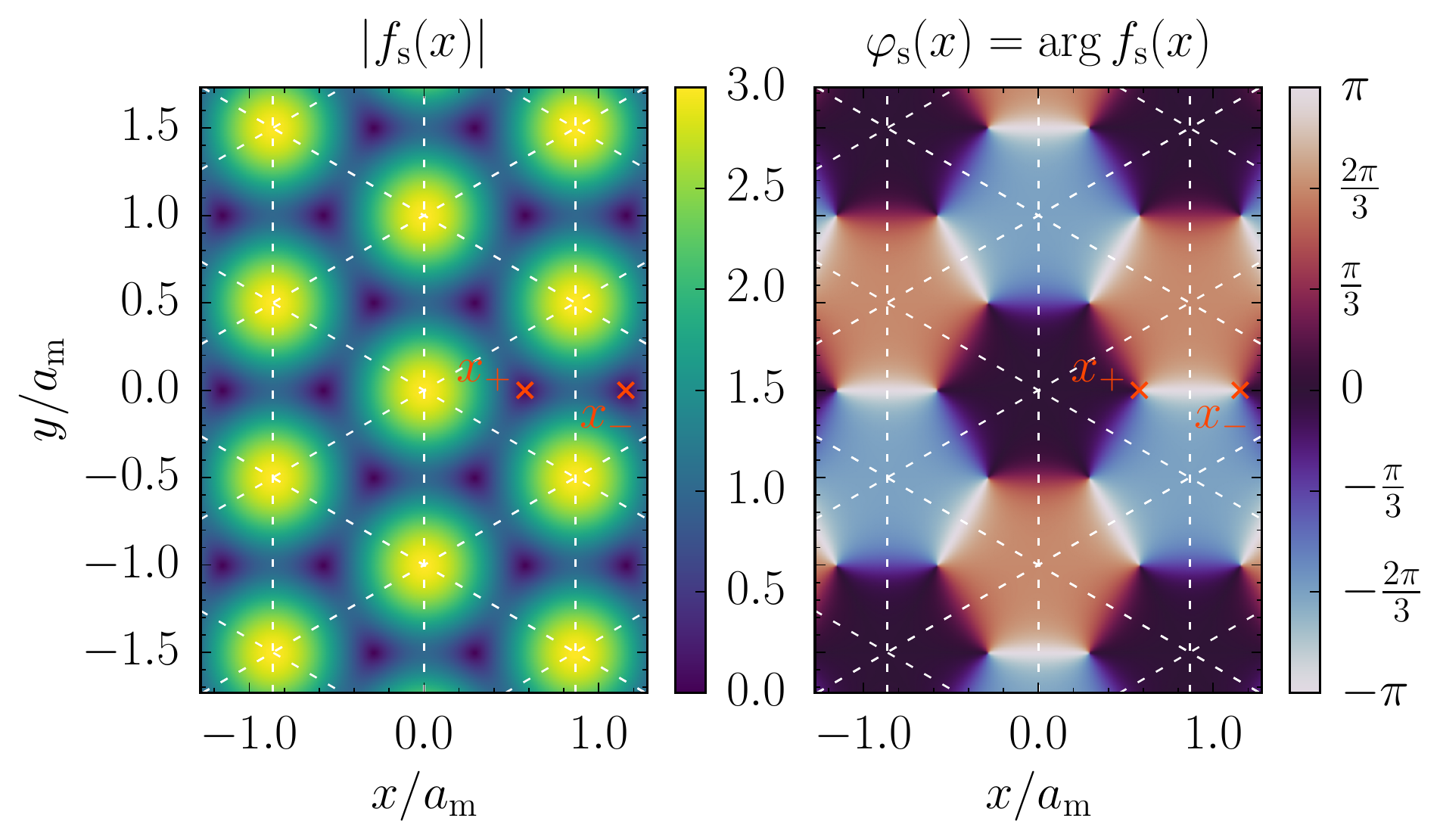}
    \caption{Left: Spatial dependence of $|f_\mathrm{s}(x)|$ (on the moiré lattice scale). At the center of each moiré triangle, the norm of $f_\mathrm{s}$ vanishes. Two singular points with positive ($x_+$) and negative ($x_-$) are marked by a red cross. Right: The phase $\phi_\mathrm{s} = \arg f_\mathrm{s}(x)$ which is seen to wind by $2 \pi$ when encircling a moiré triangle.}
    \label{fig:f}
\end{figure}

We have seen in \ref{subsec:parent_monopole} that within each layer, there are three monopoles $\Phi_{l,a}$ with $a\in \{1,2,3\}$ that carry the valley degrees of freedom and transforms in the defining representation of $\SO(3)_{\text{v}}$, while other three with $a\in \{4,5,6\}$ carry the spin degrees of freedom and transform under $\SO(3)_{\text{s}}.$
Assuming locality, hermiticity and spin-rotation invariance, the most general monopole tunneling terms are
\begin{align}
	\mathcal{L}_{1} =  \jv &\sum_{a,b=1,2,3} f_{v}^{ab} [\u_1,\u_2]\ \Phi_{1a}^\dagger(x) \Phi_{2b}(x) + \hc \nonumber\\ &+  \js\ \fs[\u_1,\u_2] \sum_{a=4,5,6}  \Phi_{1a}^\dagger(x) \Phi_{2a}(x) + \hc., 
\label{eq:interlayer}
\end{align}
In the above, $\u_l$ are spatial vectors while $x$ is a spacetime 3-vector with spatial components $\vec x$.
We will use this labeling method throughout the paper.
Since the interlayer physics should not be modified by a simultaneous translation of both layers by a lattice vector, the complex functions $\fs$ and $\fv^{ab}$ can depend only on the relative displacements between the two layers $\u\equiv \u_1-\u_2.$
One can also observe that $\fv^{ab}=\fv^a\delta_{ab}$ should be diagonal, because the translation symmetries act as diagonal matrices on the monopoles, which can be observed from the second and third columns of table \ref{tab:sym}. Only in the case $a=b$, the phase factors coming from the two layers that are gained from translations will cancel each other.

In addition, if the relative displacement field is shifted by a Bravais lattice vector of the parent triangular lattice, the interlayer physics should be invariant.
For example, under a shift of the basis vector $\a_i$ defined in Fig. \ref{fig:latbz}, $\u_1\rightarrow \u_1+\a_i$, one can compensate by $\R_l\rightarrow \R_l-\a_i$  to return to the original labelling of fields at the same locations.
Consequently, we should also make the transformations on the monopole operators as in table \ref{tab:sym}.
Considering this transformation in $\mathcal{L}_1$, we find the periodicity conditions $\fv^a[\vec u + \vec a_1] = \eu^{\iu \vec k_a \cdot \vec a_1} \fv^a[\vec u]$ and $\fs[\vec u + \vec a_1] = \eu^{\iu \vec K_1 \cdot \vec a_1} \fs[\vec u]$, which implies that the interlayer tunneling functions can be written in terms of a plane wave multiplied by a (Fourier-expanded) periodic function,
\be
\begin{split}
& \fv^a[\u]= \eu^{ \iu \k_a\cdot \u}\sum_{\Q}f_{\k_a+\Q} \eu^{\iu \Q\cdot \u},\\
& \fs[\u]= \eu^{\iu \vec K_1\cdot \u}\sum_{\Q'}f_{\vec{K}_1+\Q'} \eu^{\iu \Q'\cdot \u}.
\end{split}
\ee
with $\Q$ and $\Q'$ being the reciprocal lattice vectors and $\k_a$ the momentum of the monopole operator $\Phi_{la}^\dagger$ as defined in Eqs. \eqref{eq:k} and \eqref{eq:spin_spin}.

Next, we impose rotation symmetries. The three-fold rotation $C_3=C_6^2$ leaves invariant the spin monopoles with $a\in\{4,5,6\}$, thus relating different Fourier components in $\fs(\u)$. Keeping only the lowest harmonics, we get
\be
\fs(\u)=f_{0,s}\sum_{i=1}^3\eu^{\iu \K_i\cdot\u},
\label{eq:spin_minimal}
\ee
with $\K_i$ defined in \eqref{eq:k} and $f_{0,s}$ an arbitrary complex number for now.
A single six-fold rotation $C_6$ takes $\Phi_{la}^\dagger\rightarrow -\Phi_{la}$, giving $\fs(\u)\Phi_{1a}^\dagger(x) \Phi_{2a}(x)\rightarrow \fs(\u')\Phi_{1a}(x') \Phi_{2a}^\dagger(x')$, with $\x'=\hat{R}_{\pi/3}\x$ and $\u'=\hat{R}^{-1}_{\pi/3}\u$. ($\hat{R}_{\pi/3}$ is the counterclockwise 2d rotation matrix of angle $\pi/3$.) For this to be a symmetry, we need $\fs[\u']=\fs^*[\u]$, which, upon plugging in \eqref{eq:spin_minimal}, requires a \emph{real} $f_{0,s}=f_{0,s}^*$. 

As for the VBS monopoles with $a\in\{1,2,3\},$ rotation symmetry relates different $\fv^a$'s: Under $C_3$, $\fv^a[\u]\Phi_{1a}^\dagger\Phi_{2a}\rightarrow \fv^a[\u']\Phi_{1,a+2}^\dagger\Phi_{2,a+2}$, with $a+2$ understood as modulo $3$ and $\u'=\hat{R}^{-1}_{2\pi/3}\u$. This requires $\fv^a[\u']=\fv^{a+2}[\u]$. Keeping only the lowest harmonics closed under the symmetry, we arrive at
\be
\fv^{a}[\u]= f_{0,v}\ \eu^{\iu \k_{a}\cdot \u}.
\ee
Similarly, a single $C_6$ rotation constrains $f_{0,v}$ to be real.
The reflection symmetry does not give rise to additional constraints on the hopping amplitudes in both spin and VBS cases.
From now on, we will absorb the real constants $f_{0,s}$ and $f_{0,v}$ into the coupling constants $\jv$ and $\js$ in equation \eqref{eq:interlayer}.

In the case of a trivial stacking with $\u=0$, the monopole tunneling terms simply reduce to
\begin{equation}
	\mathcal{L}_{1}^{\text{triv}} =  \jv \sum_{a=1,2,3}  \Phi_{1a}^\dagger \Phi_{2a}  +  \js \sum_{a=4,5,6}  \Phi_{1a}^\dagger\Phi_{2a} + \hc.
\label{eq:interlayer_triv}
\end{equation}
For a rigid twist of angle $\theta$ with $\u=\theta \hat{z}\times \x$, we can write $\eu^{\iu \theta\vec K_i \cdot (\hat{z} \times \vec x)} = \eu^{\iu \theta \vec x \cdot (\vec K_i \times \hat{z})}$ and thus it becomes clear that the function $ \fs(\vec x)=\fs[\vec u(x)]$ is periodic with wavevectors vectors $\q_a=-\theta \hat{{z}} \times \vec K_a$ which lie on the corners of the lattice's moiré Brillouin zone.
In particular, we have
\begin{equation}
	\vec q_{1,3} = \frac{4 \pi \theta}{3 a_0} \left(\mp\frac{\sqrt{3}}{2},-\frac{1}{2}\right)^\top \quad \text{and} \quad \vec q_2 = \frac{4 \pi \theta}{3 a_0} \left(0, 1 \right)^\top.
\end{equation}
The $\q_a$ are hence reciprocal lattice vectors for a \emph{magnetic} moiré Brillouin zone (which is a factor 3 smaller than the  lattice's moiré Brillouin zone), and the lattice vectors for this magnetic moiré Brillouin zone read
\begin{equation} \label{eq:latvec_mmbz}
	\vec b_1^{(\mathrm{m})} = (-\sqrt{3},0)^\top \frac{a_0}{\theta} \quad \text{and} \quad \vec b_2^{(\mathrm{m})} = \left(-\frac{\sqrt{3}}{2},\frac{3}{2}\right)^\top \frac{a_0}{\theta}.
\end{equation}
For a rigid relative twist by angle $\theta$, we thus find that the interlayer monopole tunneling terms read
\be
\begin{split}
 \mathcal{L}_1= & \jv \sum_{a=1}^3 \fv^a(x) \Phi_{1a}^\dagger \Phi_{2a}+\js \fs(x)   \sum_{a=4}^6 \Phi_{1a}^\dagger \Phi_{2a}+\hc,\\
& \fv^a(x)=\eu^{\iu \q_a\cdot \x/2},\quad \fs (x)=\sum_{i=1}^3 \eu^{\iu \q_i\cdot \x}.
\end{split}
\label{eq:setup}
\ee
Note that for convenience of notation, we will define momenta in Euclidean spacetime as $q \equiv (0,\vec q)$, i.e. all interactions are at zero frequency. 
We emphasize that, while the functions $\fv^a(x)$ are \emph{pure phases}, the function $\fs(x)$ has a spatially varying magnitude and features zeroes at the the centers of the moiré triangles, as shown in Fig.~\ref{fig:f}.

\subsection{Interlayer mass couplings}
\label{subsec:mass}

In this part, we use the microscopic symmetries to constrain the possible interlayer coupling terms of bilinear masses $M_{l,\mu\nu}=\bar{\psi}_l \sigma^\mu\tau^\nu \psi_l$, with $\mu, \nu \in \{0,1,2,3\}.$ The general Lagrangian is 
\be
\mathcal{L}_2=\sum_{\mu,\nu,\rho,\sigma} w_{\mu\nu\rho\sigma}[\u_1,\u_2] M_{1,\mu\nu}(\x)M_{2,\rho\sigma}(\x).
\ee
To avoid potential confusion, we will write out all summations explicitly.
Spin rotational invariance requires $w_{\mu\nu\rho\sigma}=\delta_{\mu\rho}w_{\mu\nu\rho\sigma}$ and $w_{1\mu 1\nu}=w_{2\mu 2\nu}=w_{3\mu 3\nu}$, thus
splitting the interlayer coupling terms into two sets $\sum_i M_{1,i\mu}(\x)M_{2,i\nu}(\x)$ and $M_{1,0\mu}(\x)M_{2,0\nu}(\x)$.

Next, we impose the translation invariance similar as in the previous section: (1) Translation of the bilayer system by an arbitrary constant will not affect the interlayer physics, therefore $w_{\mu\nu\rho\sigma}[\u_1,\u_2]=w_{\mu\nu\rho\sigma}[\u]$ with $\u=\u_1-\u_2.$ (2) Based on Table~\ref{tab:mass}, under translations $T_i$, all masses are mapped to themselves up to a sign.
Therefore, we can only combine the mass terms in the two layers that obtain the same sign under these translations.
Combining with the spin rotation symmetry, we get simply $w_{\mu\nu\rho\sigma}=\delta_{\mu\rho}\delta_{\nu\sigma}w_{\mu\nu\rho\sigma}\equiv w_{\mu\nu}$. (3) We can translate only the first layer by $\a_1$.
We see that, while $M_{\mu\nu}$ with $\nu\in 2\mathbb{Z}$ transform to itself under $T_1$, the other masses obtain a minus sign.
In other words, $w_{\mu\nu}[\u]=w_{\mu\nu}[\u+\a_1]$ when $\nu$ is even, and $w_{\mu\nu}[\u]=w_{\mu\nu}[\u+2\a_1]$ when $\nu$ is odd.
\ufps{Further,} under $T_2$, we have $w_{\mu\nu}[\u]=w_{\mu\nu}[\u+\a_2]$ when $\nu\mod 3=0$, and $w_{\mu\nu}[\u]=w_{\mu\nu}[\u+2\a_2]$ when  when $\nu\mod 3=1,2.$
With this information, one can make a Fourier expansion
\be
\begin{split}
w_{\mu\nu}[\u]=\sum_{\k}{\vphantom{\sum}}' e^{i\k\cdot\u} w_{\mu\nu}(\k),
\end{split}
\label{eq:Fourier_w}
\ee
where the prime is a reminder that the summation over $\k$ has different meanings for different $\mu$ and $\nu$: In the basis of the reciprocal lattice vectors $\k=k_1\vec{b}_1+k_2\vec{b}_2$ with $\vec{b}_1=\frac{2\pi}{a_0}(1,-1/\sqrt{3})$
and
$\vec{b}_2=\frac{2\pi}{a_0}(0,2/\sqrt{3})$, $k_1\in \mathbb{Z}$ when $\nu$ is even and $k_1 \in (2\mathbb{Z}+1)/2$ when $\nu$ is odd;  $k_2\in \mathbb{Z}$ when $\nu\mod 3=0$ and $k_2 \in (2\mathbb{Z}+1)/2$ when $\nu\mod 3=1,2.$

\begin{table}[htbp]
\caption{\label{tab:mass} Transformation laws of the bilinear masses under microscopic symmetries. The second/third column represents the transformation under translation along $\vec{a}_1/\vec{a}_2$, fourth column the reflection symmetry, fifth column the six-fold rotation symmetry, and last column the time reversal symmetry. This table is adapted from table 2 in Ref. \onlinecite{song19}. }
\begin{ruledtabular}
\begin{tabular}{l|lllll}
     & $T_1$ & $T_2$ & $R$ & $C_6$ & $\mathcal{T}$ \\ \hline
     $M_{00}$ & $+$ & $+$ & $-$ & $+$ & $-$ \\ \hline 
     $M_{i0}$ & $+$ & $+$ & $+$ & $-$ & $+$ \\ \hline 
     $M_{01}$ & $-$ & $-$ & $M_{03}$ & $-M_{02}$ & $+$ \\ \hline 
     $M_{02}$ & $+$ & $-$ & $-M_{02}$ & $M_{03}$ & $+$ \\ \hline
     $M_{03}$ & $-$ & $+$ & $M_{01}$ & $M_{01}$ & $+$ \\ \hline
     $M_{i1}$ & $-$ & $-$ & $-M_{i3}$ & $M_{i2}$ & $-$ \\ \hline
     $M_{i2}$ & $+$ & $-$ & $M_{i2}$ & $-M_{i3}$ & $-$ \\ \hline
     $M_{i3}$ & $-$ & $+$ & $-M_{i1}$ & $-M_{i1}$ & $-$ \\ 
\end{tabular}
\end{ruledtabular}
\end{table}

We now examine the $w_{\mu i}$ terms under the remaining spatial symmetries.
Notice that under $C_6,$ the quartic terms $w_{\mu 1}[\u]M_{1,\mu 1}M_{2,\mu 1}\rightarrow w_{\mu 1}[\hat{O}_{C_6}^{-1}\u]M_{1,\mu 2}M_{2,\mu 2}$, and we therefore expect $w_{\mu 1}[\hat{O}_{C_6}^{-1}\u]=w_{\mu 2}[\u].$ Here $\hat{O}_{C_6}$ is the rotation matrix which sends $\a_1\rightarrow \a_2$, $\a_2\rightarrow \a_2-\a_1.$
Plugging into the Fourier series \eqref{eq:Fourier_w}, we thus arrive at $w_{\mu 1}(\k\hat{O}_{C_6}^{-1})=w_{\mu 2}(\k).$ Similarly, one can easily derive that $w_{\mu 2}(\k\hat{O}_{C_6}^{-1})=w_{\mu 3}(\k)$ and $w_{\mu 3}(\k\hat{O}_{C_6}^{-1})=w_{\mu 1}(\k).$ From the discussions of translational invariance, we know that there is no uniform $\k=(0,0)$ component of $w_{\mu i}$. The relations among the Fourier components of $w_{\mu i}$ with lowest momenta are thus
\be
\begin{split}
&\ w_{\mu 1}\left(\frac{1}{2},\frac{1}{2}\right)=w_{\mu 2}\left(0,\frac{1}{2}\right)=w_{\mu 3}\left(\frac{1}{2}, 0\right)\\
= &\ w_{\mu 1}(-\frac{1}{2},-\frac{1}{2})=w_{\mu 2}\left(0,-\frac{1}{2}\right)=w_{\mu 3}\left(-\frac{1}{2}, 0\right).
\end{split}
\ee
The reflection symmetry does not add additional constraints. 

For the $w_{\mu 0}$ terms, We have seen from spin conservation and translations, that only $w_{00} M_{1,00}M_{2,00}$ and $w_{10} \sum_i M_{1,i0}M_{2,i0}$ are allowed. This is also confirmed by the fact that $M_{\mu 0}$ can at most change by a sign under all the remaining symmetry transformations in table \ref{tab:sym}. Furthermore, similar to the  $w_{\mu i}$ case, we can derive $w_{\mu 0}[\u]=w_{\mu 0}[\hat{O}^{-1} \u]$ with $\hat{O}$ being the operator for either the $C_6$ rotation or the $M$ reflection. The Fourier components with the lowest momentum is simply the uniform piece $w_{\mu 0}(0,0)$.

Combining the discussions above and transforming back to Cartesian coordinates, we arrive at the minimal symmetry-allowed interlayer mass-mass couplings
\begin{align} \label{eq:mass_setup}
    \mathcal{L}_2 &= g_0 M_{1,00} M_{2,00} + g_1 \sum_{i=1}^3 M_{1,i0} M_{2,i0} \nonumber\\
    &+ g_2 \sum_{i=1}^3 \xi_i[\vec u] M_{1,0i} M_{2,0i} + g_3 \sum_{i=1}^3 \sum_{j=1}^3  \xi_j[\vec u] M_{1,ij} M_{2,ij} ,
\end{align}
where the functions $\xi_i[\vec u]$ are given by
\begin{equation}
    \xi_i[\vec u] = \cos(\vec{\mathsf{M}}_i \cdot \vec u),
\end{equation}
with $\vec{\mathsf{M}}_i$ denoting the $\vec{\mathsf{M}}$-points of the hexagonal Brillouin zone with $\vec{\mathsf{M}}_1 = (-\pi,-\pi/\sqrt{3})^\top/a_0$, $\vec{\mathsf{M}}_2 = (0,2 \pi/\sqrt{3})^\top/a_0$ and $\vec{\mathsf{M}}_3 = (\pi, -\pi / \sqrt{3})^\top / a_0$. 
%

%

The scaling dimension of the scalar mass $M_{00}$ in large-$N$ is found to be $\Delta_{M_{00}} = 3.08$ (to first subleading order) and the $g_0$ term is thus expected to be irrelevant \cite{chester16}.
As discussed earlier, the scaling dimension of the adjoint mass in a single layer is unclear as of now. In large-$N$, the interlayer coupling of the adjoint mass appears to be (weakly) relevant (i.e. $\Delta_M \lesssim 1.5$), but the range given in Ref.~\onlinecite{alba22} also allows for $\Delta_M \gtrsim 1.5$, potentially rendering the interlayer term irrelevant.
In the following, we thus primarily focus on the effect of the interlayer monopole tunneling terms, and revisit the effect of mass couplings in the discussion in Sec.~\ref{sec:discussion}.

{\section{Conformal perturbation theory} \label{sec:pert-theory}

We first consider the homogenously stacked system (absent of any twisting). In this case, the interlayer interaction $J$ is relevant and symmetry-allowed, and moreover the only dimensionful scale of the theory. This implies that the bilayer system is unstable in the thermodynamic limit for any infinitesimal $J$. Dimensional analysis further provides the finite-size scaling law
\begin{equation} \label{eq:j-fs-homo}
    J \sim L^{-(3-2\Delta_\Phi)},
\end{equation}
where $L$ is the linear size of the system.

To study the \emph{twisted} system's propensity towards an instability induced by the interlayer interaction, we perturbatively compute corrections to the interlayer correlator $\langle \Phi_1^\dagger(x) \Phi_2(y) \rangle$.
At long distances $\lim_{|x-y|\to\infty} \langle \Phi_1^\dagger(x) \Phi_2(y) \rangle \neq 0$ can be taken to imply long-range order.
Note that in the decoupled theory we have $\langle \Phi_1^\dagger(x) \Phi_2(y) \rangle_{\mathrm{QED}_3} = \langle \Phi_1^\dagger(x) \rangle_{\mathrm{QED}_3} \langle \Phi_2(y) \rangle_{\mathrm{QED}_3} \equiv 0$ by conformal invariance.
Upon switching on a finite (but small) $J$, this interlayer correlator no longer vanishes but will receive finite perturbative corrections which can be organized order-by-order. If these corrections are small, the physics is still controlled by the fixed point of two decoupled layers of QED$_3$.
On the other hand, the breakdown of perturbation theory (i.e. corrections are no longer small) signals an instability to some other phase/fixed point.

For explicit calculations, we find it convenient to employ a path-integral based formulation, where we denote the full action of the interacting bilayer system by $\mathcal{S} = \mathcal{S}_{\mathrm{QED}_3}^{(1)} + \mathcal{S}_{\mathrm{QED}_3}^{(2)}+\mathcal{S}_{12}$, with the associated partition function $\mathcal{Z} = \int \mathcal{D}[\{\mathcal{O}\}] \exp(-\mathcal{S})$, defining a free energy $F = -\log \mathcal{Z}$.
We will not attempt a rigorous definition of above path integral measure $\mathcal{D}[\{\mathcal{O}\}]$ which implies integrating over all operators of the conformal field theory describing the IR fixed point. In the large-$N$ limit, the measure can be rewritten in terms of the gauge-fields $a_l^\mu$ and fermions $\psi_l$ in the background of monopole configurations (which are also integrated over).
Instead, for our purposes it is sufficient to think of the monopole operators as independent fields.

In the following, we focus on the instability due to the interlayer tunneling $\js$ of the spin monopoles $\Phi_a$ with $a=4,5,6$, but an almost identical calculation holds for the VBS monopoles $\Phi_a$ with $a=1,2,3$. 
We expand the Boltzmann weight in the path integral $\eu^{-\mathcal{S}_{12}} = 1 - \mathcal{S}_{12} + \mathcal{S}^2_{12} /2 +\dots$ and denote expectation values with respect to the two copies of QED$_3$, corresponding to the two decoupled layers, by $\langle \cdot \rangle_0$.
We can then obtain the leading contribution to the interlayer correlator of the spin monopoles at large distances,
\begin{align}
    \langle \Phi_{1,a}^\dagger(d) \Phi_{2,a}(0) \rangle &\approx J \int \du^3 x \, \big[ f_\mathrm{s}^\ast(x) \nonumber\\ \times &\langle \Phi_{1,a}^\dagger(d) \Phi_{2,a}(0) \Phi_{2,b}^\dagger(x) \Phi_{1,b}(x)\rangle_{0} \big] \nonumber\\
    &\overset{|d|\to \infty}{\approx} \frac{J}{|d|^{2\Delta_\Phi}} \int \du^3 x \frac{1}{|x|^{2\Delta_\Phi}} f^\ast_\mathrm{s}(x),
\end{align}
where we have used the leading-order term of the monopole-monopole OPE (which corresponds to the two-point function \eqref{eq:twopoint}) in the two layers.
Recalling $f_\mathrm{s}(x) = \sum_{i=1}^3 \eu^{\iu \vec q_i \cdot \vec x}$, the integral $\int\du^3 x \, |x|^{-2\Delta_\Phi} f^\ast_\mathrm{s}(x) \sim |q|^{2 \Delta_\Phi - 3}$ converges, where the wavevector $|q|$ is proportional to the inverse moir\'e lattice constant, $|q| \sim a_\mathrm{m}^{-1}$. Hence, at at leading order we have
\begin{equation} \label{eq:leading-pert-theory}
     \langle \Phi_{1,a}^\dagger(d) \Phi_{2,a}(0) \rangle \overset{|d|\to\infty}{\sim} \frac{J |q|^{-(3-2\Delta_\Phi)}}{|d|^{2\Delta_\Phi}}.
\end{equation}
Evaluating the next term in the perturbative expansion, quadratic in $\mathcal{S}_{12}$, necessitates evaluating three-point functions of the form $\langle \Phi^\dagger \Phi \Phi \rangle_{\mathrm{QED}_3}$ (and hermitian conjugates) in each layer. Given that monopole operators are charged under the (emergent) $\Uone_\mathrm{top}$ symmetry of the \qed{} fixed-point theory, it becomes clear that any correlation function of an \emph{odd} number of monopole operators needs to vanish identically.
The first (subleading) correction to \eqref{eq:leading-pert-theory} therefore occurs at third order and involves four-point correlation functions in each layer.
Four-point correlation functions can be decomposed by successively applying the OPE. In principle, any choice of order is equivalent as the OPE is convergent.
However, as we are truncating the OPE beyond leading order, these different choices become inequivalent. Each OPE channel corresponds to the most divergent contributions from respective regions of configuration space where the operators are ``close''.

In order to investigate the stability of the system at hand, we should therefore find the OPE channel with the strongest IR divergence.
We find \footnote{Note that there is a more divergent term which corresponds to using the OPE $\Phi_1(d)^\dagger \Phi_1(x) \sim |d-x|^{-2\Delta_\Phi}$ \emph{and} $\Phi_2(x)^\dagger \Phi_2(0) \sim |x|^{-2\Delta_\Phi}$, implying that $x$ is simultaneously close to $0$ and $d$, which contradicts our assumption of $|d|\to \infty$.}
\begin{widetext}
\begin{align}
    &\delta \langle \Phi_{1,a}^\dagger(d) \Phi_{2,a}(0) \rangle \sim J^3 \int \du^3 x \, \du^3 y \, \du^3 z \Big[ f^\ast_\mathrm{s}(x) f_\mathrm{s}(y) f_\mathrm{s}^\ast(z) \langle \Phi^\dagger_{1,a}(d) \Phi_{2,a}(0) \Phi_{1,b}(x) \Phi_{2,b}^\dagger(x) \Phi_{1,c}^\dagger(y) \Phi_{2,c}(y) \Phi_{1,d}(z) \Phi_{2,d}^\dagger(z) \rangle_{0} \Big] \\
    &\sim J^3 \int \du^3 x \, \du^3 y \, \du^3 z \,  \frac{\sum_{i,j,k} \eu^{-\iu (q_i \cdot x - q_j \cdot y + q_k \cdot z)}}{|d-x|^{2\Delta_\Phi} |z|^{2\Delta_\Phi} |y-z|^{2\Delta_\Phi} |x-y|^{2\Delta_\Phi} } \overset{|d|\to\infty}{\sim} \frac{J^3 |q|^{2\Delta_\Phi-3}}{|d|^{2\Delta_\Phi}} \int \du^3 y' \, \du^3 z \frac{\sum_{i,j,k} \eu^{-\iu(q_i - q_j) \cdot y'} \eu^{-\iu (q_i-q_j +q_k) \cdot z}}{|y'|^{2 \Delta_\Phi} |z|^{2\Delta_\Phi}}. \nonumber
\end{align}
\end{widetext}
In the last step, we have substituted $x' = x-y$ and performed the $x'$-integration and similarly substituted $y'=y-z$.
We note that for the summands with $i=j$ the $y'$-integral reduces to $\int \du^3 y' |y'|^{-2\Delta_\Phi} \sim L^{3-2\Delta_\Phi}$, which has an IR-divergence that we regularize with the system size $L$. On the other hand, the $z$-integration is regular (the potential IR divergence is cut off by the oscillatory exponential for all index combination $i,j,k$), and we therefore arrive at
\begin{equation}
    \delta \langle \Phi_{1,a}^\dagger(d) \Phi_{2,a}(0) \overset{|d|\to \infty}{\sim} \frac{J^3 |q|^{2 \cdot (2 \Delta_\Phi -3)} L^{3-2\Delta_\Phi} }{|d|^{2\Delta_\Phi}}. 
\end{equation}
As discussed above, if this subleading correction is no longer small compared to the leading order result \eqref{eq:leading-pert-theory}, perturbation theory breaks down.
We can therefore obtain the scaling relation which determines the critical point by asking when the ratio of subleading correction to the leading order term is of order unity,
\begin{equation}
    \frac{\delta \langle \Phi_{1,a}^\dagger(d) \Phi_{2,a}(0) \rangle}{ \langle \Phi_{1,a}^\dagger(d) \Phi_{2,a}(0) \rangle} \overset{!}{\sim} 1 \quad \Rightarrow \quad J^2 |q|^{2\Delta_\Phi - 3} L^{3-2\Delta_\Phi} \sim 1.
\end{equation}
Importantly, we find that this condition for the instability still depends on the system size $L$, which was required to regularize IR divergence.
Rewriting above result as
\begin{equation} \label{eq:j-fs-twist}
    J \sim \left(a_\mathrm{m} L\right)^{- \left(\frac{3}{2} -\Delta_\Phi \right)}
\end{equation}
makes clear that in the thermodynamic limit (i.e. $L\to \infty$ with $a_\mathrm{m}$ fixed) an infinitesimal $J$ is sufficient to induce an instability, implying that the bilayer system is unstable upon twisting, as in the trivially stacked case (recall $3/2 > \Delta_\Phi \approx 1.05$ as discussed in Sec.~\ref{sec:data_OPE}).

Crucially however, comparing with the finite-size scaling law for the homogenously stacked system in \eqref{eq:j-fs-homo}, we find that the critical $J$ in \eqref{eq:j-fs-twist} scales slower as a function of $L$, which implies that there is a parametrically large region (as a function of $L$, compared to the homogenously stacked case) in which the twisted system remains stable.

Consequently, our perturbative analysis implies that while an instability remains at finite twist angles, we find that for any $\theta \neq 0$ the instability is \emph{softened} in the sense that the interlayer interaction has effectively become less relevant.

}

\section{Variational conformal mean-field theory}
\label{sec:mf}

{While the breakdown of perturbation theory at the (finite-size) critical point \eqref{eq:j-fs-twist} suggests that an instability occurs, the perturbative approach is not capable of describing the nature of the resulting phase.
Given that monopole operators transform as order parameters for Néel/VBS order on the triangular lattice, it appears plausible that the interlayer interaction \eqref{eq:setup} leads to some ordered phase.
We therefore employ} mean-field theory, which consists in approximating the monopole-antimonopole interaction by monopole operators coupling to a mean-field which is determined self-consistently.
Physically, the mean fields being finite signals that monopole operators have condensed and thus the onset of VBS or magnetic order.

Commonly, the virtue of mean-field theory is that it replaces the task of solving an interacting problem with the solution of a non-interacting problem (which can be done exactly) and finding appropriate self-consistent parameters.
In the model at hand, we emphasize that the mean-field approximation still requires us to solve \qed{} with monopole operators coupled to some classical background field.
As mentioned above, even in the absence of such background field, \qed{} is not exactly solvable and believed to be described by a strongly-interacting fixed point.
However, as we show below, the conformal structure of said fixed point places strong constraints on correlation functions and scaling behaviors, which in turn allow us to evaluate certain observables in \qed{} as functions of the classical background field in controlled limits.

{As defined in the previous section, the system's free energy is given by $F=-\log \mathcal{Z}$ with the partition function $\mathcal{Z} = \int \mathcal{D}[\{\mathcal{O}\}] \exp(-\mathcal{S})$ and the action $\mathcal{S} = \mathcal{S}_{\mathrm{QED}_3}^{(1)} + \mathcal{S}_{\mathrm{QED}_3}^{(2)}+\mathcal{S}_{12}$.}
Using Jensen's inequality \cite{feyn98}, it follows that $F$ obeys the Bogoliubov-Gibbs-Feynman inequality 
\begin{equation} \label{eq:bgf-ineq}
	F \leq F_\mathrm{mf} + \langle \mathcal{S} - \mathcal{S}_\mathrm{mf} \rangle_\mathrm{mf} \equiv F_\mathrm{var},
\end{equation}
where $\mathcal{S}_\mathrm{mf}[\bm{h}_1,\bm{h}_2] = \sum_{l=1,2} \mathcal{S}_{\mathrm{QED}_3}^{(l)} + \mathcal{S}^{(l)}_h[\bm{h}_l]$ is the mean-field action corresponding to two decoupled DSL with classical fields $\bm{h}_l$ (``mean fields'') coupling to the monopole operators, $\mathcal{S}_h^{(l)}[\bm{h}_l] = \int \du^3x \left[-{h}_l^{a\ast}(x) \Phi_{l,a}(x) + \mathrm{h.c.}\right]$. $F\mf$ is the corresponding mean-field free energy. The $\bm{h}_l$ are subject to self-consistency equations upon minimizing $F_\mathrm{var}$, as shown below. 

We use brackets to indicate that the mean-field partition functions and expectation values are (via the action $\mathcal{S}_\mathrm{mf}$ in the Boltzmann factor) functionals of the fields $h_{l}^a(x)$, with $l=1,2$ denoting the layer, and $a=1,\dots,6$ a $\SOsix$ index.
Defining the single-layer partition function
\begin{equation}
	\mathcal{Z}_\mathrm{mf} [\bm{h}_l] = \int \mathcal{D}[\{\mathcal{O}\}] \eu^{-\mathcal{S}_{\mathrm{QED}_3}^{(l)} - \mathcal{S}_h[\bm{h}_l]},
\end{equation}
the partition function associated with $\mathcal{S}_\mathrm{mf}$ factorizes and thus the mean-field free energy $F_\mathrm{mf} = -\log \mathcal{Z}_\mathrm{mf} \equiv - \log \mathcal{Z}_\mathrm{mf}[\bm{h}_1] - \log \mathcal{Z}_\mathrm{mf}[\bm{h}_2]$.
We thus rewrite \eqref{eq:bgf-ineq} as
\begin{equation} 
\label{eq:bgf-expanded}
\begin{split}
	F \leq - & \log \mathcal{Z}_\mathrm{mf}[\bm{h}_1] - \log \mathcal{Z}_\mathrm{mf}[\bm{h}_2] + \langle\mathcal{S}_{12}\rangle_\mathrm{mf}[\bm{h}_1,\bm{h}_2] \\
	& - \sum_{l=1,2}\langle S_h^{(l)} \rangle_\mathrm{mf} [\bm{h}_l].
\end{split}
\end{equation}
Note that the expectation values of the interlayer interactions factorize due to the linearity of $\mathcal{S}_\mathrm{var}$ such that $\langle \Phi_{1,a}^\dagger \Phi_{2,b} \rangle_\mathrm{mf}[\bm{h}_1,\bm{h}_2] = \langle \Phi_{1,a}^\dagger \rangle[\bm{h}_1] \langle \Phi_{2,b} \rangle_\mathrm{mf}[\bm{h}_2]$.

Next, we seek to minimize the right-hand-side of \eqref{eq:bgf-expanded} with respect to the functions $\bm{h}_l$.
To this end, we first note that
$\delta/\delta h_{l}^{a\ast}(x) \log \mathcal{Z}_0[h_{l'}] = \langle \Phi_{l,a}(x) \rangle_\mathrm{mf} \delta_{l,l'}$, which embodies that $\langle \Phi \rangle$ and $h^\ast$ are conjugate variables by construction, and further
\begin{equation}
    \frac{\delta}{\delta h_{l'}^{b\ast}(y)} \langle \Phi_{l,a}^\dagger(x) \rangle_\mathrm{mf} = \langle \Phi^\dagger_{l,a}(x) \Phi_{l,b}(y) \rangle_\mathrm{mf} \delta_{l,l'},
\end{equation}
and similarly for $\langle \Phi_{l,a}(x) \rangle_\mathrm{mf}$, which yields the anomalous monopole correlation function.
After some manipulations, the saddle-point condition $\delta F_\mathrm{var}/\delta h_{1}^{b\ast}(y) = 0$ is rewritten as
\begin{widetext}
\begin{equation} \label{eq:saddle-point-int}
	0 = \int \du^3 x  \sum_{a=1}^6 \left[ \big( J_a  f_a(x)\langle \Phi_{2,a}(x)\rangle\mf + h_{1}^a(x) \big) \langle \Phi_{1,a}^{\dagger}(x)\Phi_{1,b}(y) \rangle_\mathrm{mf} + \big( J_a  f_a^\ast(x)\langle \Phi_{2,a}(x)\rangle\mf + h_{1}^{a\ast}(x) \big) \langle \Phi_{1,a}(x)\Phi_{1,b}(y) \rangle_\mathrm{mf}  \right],
\end{equation}
\end{widetext}
for all $x$, and $b=1,\dots,6$.
A similar equation is obtained from $\delta F_\mathrm{var}/\delta h_{2}^{b\ast}(y) = 0$. Here, we have employed $J_{1,2,3} \equiv \jv$ and $J_{4,5,6} \equiv \js$, as well as $f_{1,2,3} = \fv^{1,2,3}$ and $f_{4,5,6} \equiv \fs$.
While \eqref{eq:saddle-point-int} is hard to solve directly, a \emph{sufficient} condition for the integral to vanish is given by
\begin{subequations}
\begin{align} 
    h_{1}^a(x) &= - \jv \fv^a(x) \langle \Phi_{2,a} \rangle_\mathrm{mf}[\bm{h}_2] \\
    h_{2}^a(x) &= - \jv \fv^{a\ast}(x) \langle \Phi_{1,a} \rangle_\mathrm{mf}[\bm{h}_1]
\end{align}
\label{eq:self-con-VBS}
\end{subequations}
for $a=1,2,3$ \emph{and}
\begin{subequations}
\begin{align} 
    h_{1}^a(x) &= - \js \fs(x) \langle \Phi_{2,a} \rangle_\mathrm{mf}[\bm{h}_2] \\
    h_{2}^a(x) &= - \js \fs^\ast(x) \langle \Phi_{1,a} \rangle_\mathrm{mf}[\bm{h}_1] 
\end{align}
\label{eq:self-con-S}
\end{subequations}
for $a=4,5,6$.

The expectation values $\langle \cdot \rangle_\mathrm{mf}$ implicitly depend on $\bm{h}_1$ and $\bm{h}_2$ (note that the dependence on $\bm{h}_2$ drops out when the expectation value is taken of operators in layer $1$ only, and vice versa).
Hence, Eqs.~\eqref{eq:self-con-VBS} and \eqref{eq:self-con-S} constitute a set of self-consistency equations for the mean fields $\bm{h}_1$ and $\bm{h}_2$.
We note that these become a \emph{necessary condition} for \eqref{eq:saddle-point-int} to vanish \emph{iff} the normal and anomalous monopole correlation functions are translationally invariant, $\langle \Phi_{1,a}^{(\dagger)}(x)\Phi_{1,b}(y) \rangle \equiv \langle \Phi_{1,a}^{(\dagger)}(x-y)\Phi_{1,b}(0) \rangle$.
Then \eqref{eq:saddle-point-int} can be rewritten as a convolution which vanishes if either Kernel or argument are zero.

{\section{Weak-coupling solution of mean-field theory}}
\label{sec:perturbativeMFT}

We consider a scaling transformation to a length scale set by the Moiré lattice scale $a_\mathrm{m} \sim 1/|q| \sim a_0/ \theta$.
Performing such transformation explicitly in \eqref{eq:setup}, we find that the interlayer interaction is multiplied by a dimensionless parameter, $J / |q|^{3-2 \Delta \Phi}$. 
This parameter being small corresponds to interlayer couplings which are small compared to (fast) modulations of the interlayer tunneling amplitude due to Moiré modulations, $J \ll a_\mathrm{m}^{2 \Delta_\Phi -3}$.
Using \eqref{eq:self-con-VBS} and \eqref{eq:self-con-S}, this further implies that the mean fields can be taken to be small as $|\bm{h}_{1,2}| \sim J q^{\Delta_\Phi} \ll a_\mathrm{m}^{\Delta_\Phi-3}$.
In this limit, we can evaluate expectation values $\langle \cdot \rangle\mf[\bm{h}_l]$ perturbatively order-by-order in the background field $\bm{h}_l$, allowing us to write down and solve \emph{linearized} self-consistency equations.

\subsection{Linearized self-consistency equations} 

We start by evaluating the expectation value $\bk{\Phi_{la}}\mf$.
Expanding $\eu^{-\mathcal{S}\mf}=\eu^{-\mathcal{S}\0}(1-\mathcal{S}_h+\mathcal{S}_h^2/2+\cdots)$, we have 
\be
\begin{split}
\bk{\Phi_{la}(x)}\mf=\ & \mathcal{Z} \mf^{-1} \int \mathcal{D}[\{\mathcal{O}\}]\  \Phi_{la}(x) \eu^{-\mathcal{S}\mf} \\
=\ &  \int \du^3y \ \frac{h_{l}^{a}(y)}{|x-y|^{2\Delta_\Phi}}  +O(h^3),\\
\end{split}
\ee
where we have used the OPE of the monopoles \eqref{eq:OPE-def} at leading order.
Note that this result is equivalently obtained by integrating out all \qed{} degrees of freedom perturbatively to obtain $F\mf[\bm{h}_1,\bm{h}_2]$ at quadratic order in $\bm{h}_1$,$\bm{h}_2$, and then using $\langle\Phi_{la}\rangle\mf = - \delta F\mf /\delta h_{l}^{a\ast}$.
 
Plugging this into the mean-field equations \eqref{eq:self-con-VBS} and \eqref{eq:self-con-S}, we arrive at the relationship between the effective fields in the two layers 
\be
\begin{split}
& h_1^a(x)=-J_a f_a(x)  \int d^3y \ \frac{h_{2}^{a}(y)}{|x-y|^{2\Delta_\Phi}},\\
& h_2^a(x)=-J_a f^{*}_a(x)  \int d^3y \ \frac{h_{1}^{a}(y)}{|x-y|^{2\Delta_\Phi}},
\end{split}
\label{eq:coupled_h}
\ee
where again $J_a=\jv$, $f_a=\fv^a$ for $a\in\{1,2,3\}$ and $J_a=\js$, $f_a=f_s$ for $a\in\{4,5,6\}$. There is no implicit summation over $a$. Eliminating $h_2^a$ from above, we arrive at
\be
\begin{split}
h_1^a(x)=J_a^2 f_{a}(x) & \int d^3z\ h_1^a(z)\int d^3y\ f_{a}^{*}(y) \\
& \times  \frac{1}{|y-z|^{2\Delta_\Phi}} \frac{1}{|x-y|^{2\Delta_\Phi}}.
\end{split}
\label{eq:target}
\ee
There is also a similar equation for $h_2^a(x)$.
In the next two subsections, we will discuss the cases of spin and VBS monopoles, separately. 
From \eqref{eq:coupled_h} and \eqref{eq:target} it is clear that the perturbative approach employed here does not fix the magnitude of the mean fields $h_1^a$ and $h_2^a$.
Rather, they determine for which critical $\js,\jv$ the self-consistency equations admit non-trivial solutions with $h_1^a, h_2^a \neq 0$ and the symmetry of the mean fields.

\subsection{Spin monopoles}

\subsubsection{Solution of self-consistency equations}

Given the periodicity of the Moiré pattern, we Fourier-expand the mean fields as ${\bm{h}}_1(x) = \sum_{Q} \tilde{\bm{h}}_1(Q) \eu^{\iu Q \cdot x}$ on both sides of \eqref{eq:target}, with some to-be-determined spacetime momenta $Q$.
Note that because of instantaneous nature of the interaction (i.e. $q_i = (0,\vec q_i)^\top$ in $\fs(x)$ and $\fv(x)$ as given in \eqref{eq:setup}), we can immediately write $q_i \cdot x = \vec q_i \cdot \vec x$
We further expect $Q$ to be some linear combination of the $q_i$ and thus $\tilde{\bvec{h}}_l(Q) \equiv \tilde{\bvec{h}}_l(0,\vec Q) \equiv \tilde{\bvec{h}}_l(\vec Q)$ to denote the Fourier coefficients of $\bvec{h}_l(x)$.
We hence find
\be
\begin{split}
\sum_{\Q'}\tilde{\bm{h}}_1(\Q')\eu^{\iu \Q'\cdot \x}= & \js^2\ \fs(\x) \sum_i \int d^3y\frac{\eu^{-\iu \q_i\cdot \vec{y}}}{|y-x|^{2\Delta_\Phi}} \\
& \times \sum_{\Q} \tilde{\bm{h}}_1(\Q) \int d^3z\  \frac{\eu^{\iu \Q \cdot \vec{z}} }{|z-y|^{2\Delta_\Phi}}
\end{split}
\label{eq:consistency1}
\ee
Upon changing of variables $y-x\rightarrow y$ in the first integral and $z-y\rightarrow z$ in the second integral, the equation simplifies to
\be
\begin{split}
\sum_{\Q'}\tilde{\bm{h}}_1(\Q')\eu^{\iu \Q'\cdot \x}= & \js^2\ \fs(\x) \sum_{\Q}  \tilde{\bm{h}}_1(\Q)\sum_i \eu^{\iu(\Q-\q_i)\cdot \x}  \\
& \times  \int d^3y\frac{\eu^{\iu(\Q-\q_i)\cdot \vec{y}}}{|y|^{2\Delta_\Phi}} \int d^3z\  \frac{\eu^{\iu \Q\cdot \vec{z}}}{|z|^{2\Delta_\Phi}}.
\end{split}
\label{eq:consistency2}
\ee
Integrals of the form $\int \du^3 x\ \eu^{\iu \k\cdot \x}/|x|^{2\Delta_\Phi}$ diverge when $\vec{k}=0$. In the equation above, only one of the two integrals can possibly diverge for any given wavevector $\Q$, corresponding to cases (i) $\Q=\q_i$ or (ii) $\Q=0$, respectively.
When $\Delta_\Phi$ is smaller than $3/2$, i.e. the monopole tunneling term is a relevant perturbation to the $\js = \jv = 0$ fixed point, the divergence is in the infrared (IR) limit.
As discussed in Sec.~\ref{sec:data_OPE}, large-$N$ calculations and a recent conformal bootstrap study points $\Delta_\Phi \simeq 1.02 < 3/2$, such that the interlayer term can be assumed to be strongly relevant, and we have indeed an IR divergence at hand.

To regulate above divergence, we introduce an IR cutoff $L> |x-y|$ to bound the maximal separation of two monopole operators (at coordinates $x$ and $y$) from above.
We emphasize this is a cutoff for the Euclidean space-time integrals.
The length scale set by the cutoff $L$ can therefore be interpreted the system's linear size (i.e. its spatial extent) and simultaneously as an inverse temperature $\beta \sim 1/L$ (for the imaginary time direction).
The limit $L\to \infty$, which is ultimately of interest to us,  thus corresponds to the zero-temperature thermodynamic limit of an infinitely large system.

Having introduced above cutoff scheme, the two singular contributions are
\begin{widetext}
\be
\begin{split}
\sum_{\Q'}\tilde{\bm{h}}_1(\Q')\eu^{\iu \Q'\cdot \x}& =\js^2 \fs(\x)\sum_i \left\{\eu^{-\iu \q_i\cdot \x}\tilde{\bm{h}}_1(0)\int \du^3y \frac{e^{-i\q_i\cdot \vec{y}} }{|y|^{2\Delta_\Phi}} \int \du^3z \frac{1}{|z|^{2\Delta_\Phi}} + \tilde{\bm{h}}_1(\q_i) \int \du^3y \frac{1}{|y|^{2\Delta_\Phi}} \int d^3z \frac{\eu^{\iu \q_i\cdot \vec{z}}}{|z|^{2\Delta_\Phi}} +\cdots\right\}\\
& =c \js^2 \left|\frac{q}{L}\right|^{2\Delta_\Phi-3} \sum_{i,j} [\tilde{\bm{h}}_1(0)\eu^{\iu(\q_j-\q_i)\cdot x}+\tilde{\bm{h}}_1(\q_i)\eu^{\iu \q_j\cdot x}]+\cdots,\label{eq:fourier-expansion}
\end{split}
\ee
\end{widetext}
with $c=16\pi^2 \Gamma(2-2\Delta_\Phi)\sin(\Delta_\Phi\pi)/(3-2\Delta_\Phi)$ a constant, $|q|=4\pi\theta/3a_0$ from the discussions in section \ref{subsec:bootstrap}, and $\cdots$ represents the contributions from other Fourier components that are regular.
Comparing the spatial dependence on the two sides, $\vec{Q}'$ should belong to the set $\{\q_i-\q_j, \q_i\}$ for arbitrary $i,j$. In the following, we are interested in the long-wavelength modulations of the mean fields and hence drop all oscillatory terms with wavevectors outside the first moiré Brillouin zone (note that $(\q_i - \q_j) \in \mathrm{1^{st}}$ MBZ iff $i=j$). 

Since the equation holds for arbitrary $x$, we can equate the Fourier components directly.
For both $\tilde{\bm{h}}_1(\Q=0)$ and $\tilde{\bm{h}}_1(\Q=\q_i)$, this leads to
\be
\tilde{\bm{h}}_1(\Q)=3c\js^2 \tilde{\bm{h}}_1(\Q)\left|\frac{q}{L}\right|^{2\Delta_\Phi-3}.
\label{eq:constraint_J}
\ee
Note that $\tilde{\bm{h}}_1(\vec{Q})=0$ trivially satisfies this equation.
A nontrivial solution exist for $3 c J_s^2 |q/L|^{2 \Delta_\Phi-3}\equiv1$, such that \eqref{eq:constraint_J} becomes a constraint for $\js$ leading to the critical interlayer tunneling strength with scaling
\be
\js\sim  (a_\mathrm{m} L)^{\Delta_\Phi-3/2},
\label{eq:js}
\ee
with $a_\mathrm{m} \sim a_0/\theta\ll L$ the moiré lattice constant.
{This is precisely the same critical scaling for the twisted system as obtained via perturbation theory in the interlayer coupling (\emph{without} any additional (mean-field) approximation) in Eq.~\eqref{eq:j-fs-twist}.
The fact that our mean-field treatment recovers the correct critical scaling for the instability is a non-trivial cross-check and provides confidence for the reliability of our mean-field theory.}

While the perturbative approach does not fully determine the order parameter, we can extract its symmetry properties. 
Neglecting higher-wavevector oscillations, Eqs.~ \eqref{eq:fourier-expansion} and \eqref{eq:constraint_J} suggest the minimal form of $h_1(x)$ as
\be
\bm{h}_1(\x)\approx \tilde{\bm{h}}_1(0)+\tilde{\bm{h}}_1(q)\fs(\x),
\ee
where $q=|q_i|$ is independent of the index $i$, and we emphasize that the magnitude of the Fourier coefficients $\tilde{h}(0)$, $\tilde{h}(q)$ is arbitrary:
Due to the linearity of the self-consistency equations at weak coupling, any linear superposition of the two degenerate solutions with finite $\tilde{h}_1(0) \neq 0$ and $\tilde{h}_1(q) \neq 0$ is also a solution to the self-consistency equations.   Similarly, if we eliminate $h_1$ from equation \eqref{eq:coupled_h}, we would get
\be
\bm{h}_2(\x)\approx \tilde{\bm{h}}_2(0) +\tilde{\bm{h}}_2(-q)f_s^*(\x).
\ee
These four Fourier components in the two layers are related to each other through \eqref{eq:coupled_h}:
\be
\begin{split}
& \tilde{\bm{h}}_1(q)=-A \tilde{\bm{h}}_2(0),\quad \tilde{\bm{h}}_2(-q)=-A \tilde{\bm{h}}_1(0),\\
& A\equiv \frac{4\pi \js}{3-2\Delta_\Phi} L^{3-2\Delta_\Phi}\sim (L/a_\mathrm{m})^{-\Delta_\Phi+3/2},\\
\end{split}
\label{eq:A}
\ee
where we have used the critical condition \eqref{eq:js} in the second line. Therefore, when $\js$ is positive, generically we have $A\gg 1$, and the spatial modulating pieces in $\bm{h}_l(x)$ dominate.

The relation between $\tilde{\bm{h}}_1(0)$ and $\tilde{\bm{h}}_2(0)$ can be further obtained by going to quartic order in perturbation (see appendix \ref{app:spin} for details), giving rise to $\tilde{\bm{h}}_1(0)= r \tilde{\bm{h}}_2(0)$ with $r$ being a constant phase.
This $r$ can then be fixed by noticing that in the trivial stacking limit $|q|\rightarrow 0$, when $\js<0$ (or $\js>0$), we expect the Néel order parameters to align (or anti-align).
We can therefore determine $r=- \mathrm{sign}(\js).$

Combining above results with the mean field equations \eqref{eq:self-con-S} , we finally arrive at the following spatial dependence of the spin monopole expectation values
\be
\begin{split}
& \bk{\bm{\Phi}_{1}(\x)}\mf \approx -\frac{A}{|\js|}\tilde{\bm{h}}(0) -\frac{\fs(x)}{3\js }\tilde{\bm{h}}(0),\\
& \bk{\bm{\Phi}_{2}(\x)}\mf \approx  \frac{A}{\js}\tilde{\bm{h}}(0)+\frac{\fs^*(x)}{ 3|\js|}\tilde{\bm{h}}(0),\\
\end{split}
\label{eq:weak_Phi}
\ee
where for consistency, we  again only keep wavevectors in the first moiré Brillouin zone (this implies that $\fs^\ast(x) \fs(x) \approx 3$).
We have added back the flavor degrees of freedom and used boldface to label the three-dimensional vectors in the $\SO(3)_{\text{s}}$ space. 
Since $A\gg 1$, we observe that the first terms in the two equations above dominate, and there are some corrections with modulations of moiré scale.

\subsubsection{Lifting of \emph{global} $\mathrm{U}(3)$ degeneracy at quartic order}

The above weak-coupling analysis determines the ordering wavevectors as well as relative phase factors of the mean fields (or, equivalently, of the order parameter) in the two layers.
However, so far our analysis has not determined the form of the Fourier coefficients $\tilde{\bm{h}}_l(0)$ and $\tilde{\bm{h}}_{l}(\pm q)$ which are understood to be three-dimensional vectors.
Indeed, the self-consistency equations at quadratic order \eqref{eq:target}, or equivalently \eqref{eq:fourier-expansion}, which lie at the heart of our analysis, show that there is a \emph{global} $\mathrm{U}(3)$ invariance of rotating $\bm{h}_{1,2}(x) \mapsto g \bm{h}_{1,2}(x)$ with $g \in \mathrm{U}(3)$ which does not correspond to a \emph{physical} symmetry operation.
This redundancy is understood to be an artifact of the quadratic approximation:
For a single layer, the mean-field free energy, after perturbatively integrating out \qed{} degrees of freedom, is given by (note that we omit the layer index)
\begin{equation} \label{eq:pert-quadratic}
	F\mf[\bvec{h}] = - \log \mathcal{Z}_\mathrm{mf}[0] - \int \du^3 x \, \du^3 y \frac{\bvec{h}^\ast(x) \cdot \bvec{h}(y)}{|x-y|^{2 \Delta_\Phi}} + O(h^3),
\end{equation}
which is readily seen to be invariant under global $\mathrm{U}(6)$ transformations which reduce to $\mathrm{U}(3) \subset \mathrm{U}(6)$ when considering the spin sector with $h^1 = h^2 = h^3 \equiv 0$, where the upper indices are the $\SOsix$ flavor indices.
While the form of the quadratic term \eqref{eq:pert-quadratic} is unique and mandated by symmetry, an analysis of higher-order $\SO(6)$-invariant tensor structures reveals that at quartic order, this accidental degeneracy is broken.
Note that there are no odd-order terms in the perturbative expansion of $F\mf$ by $\Uone_\mathrm{top}$ and $\SO(6)$ symmetry.

Expanding perturbatively in $h$, the mean-field free energy (for a single layer) is written as $F\mf[\bvec{h}] = - \log \mathcal{Z}_\mathrm{mf} - \langle \mathcal{S}_h^2 \rangle_0/2 - \left( \langle \mathcal{S}_h^4 \rangle_0 - 3 \langle \mathcal{S}_h^2 \rangle_0^2 \right)/4!+ \dots$, where $\langle \cdot \rangle_0$ denotes evaluating expectation values in the unperturbed compact \qed{} theory.
The quartic term necessitates the evaluation of
\begin{equation} \label{eq:quartic-pure}
	\langle \mathcal{S}_h^4 \rangle_0 = \int \du^3 x_1 \dots \du^3 x_4 \, \langle \prod_{i=1}^4  (h^{a_i\ast}(x_i) \Phi_{a_i}(x_i) + \hc) \rangle_0,
\end{equation}
which requires knowledge of the four-point function of monopole operators $\langle \Phi_a^{s_1}(x_1) \Phi_b^{s_2}(x_2) \Phi_c^{s_3}(x_3) \Phi_d^{s_4}(w) \rangle_0$ (here $s_i = \pm$ and $\Phi^{+}\equiv \Phi$, $\Phi^{-}\equiv \Phi^\dagger$ should be understood as monopole and antimonopole, respectively).
In contrast to two- and three-point functions, conformal symmetry \emph{does not} fully determine four-point functions.
Rather, they can be written in terms of so-called \emph{conformal blocks}, which are functions of conformally invariant parameters and obey certain associativity relations which are used in the conformal bootstrap approach \cite{he21,alba22}.

In 2+1 dimensions, however, no closed form for the conformal blocks exist, and we instead follow an approximate strategy.

Firstly, we use the fact that the finite-wavevector components of $\bm{h}_l(x)$ can be related to the constant components in the respective other layer, $\tilde{\bm{h}}_l(0)$, such that it is sufficient to consider uniform $\bm{h}_l(x) = \mathrm{const.}$ in \eqref{eq:quartic-pure}.
By $\Uone_\mathrm{top}$ symmetry, only expectation values with zero net topological charge (i.e. containing two monopoles and two antimonopole operators) can be finite.
We then posit that a dominant contribution to the integral in \eqref{eq:quartic-pure} is given by configurations where operator insertions are \emph{close} to each other.
Concretely, we consider $\langle \Phi_a^\dagger(x_1) \Phi_b(x_2) \Phi_c^\dagger(x_3) \Phi_d(x_4)\rangle_0$ (the other terms follow by index permutations) and take $x_1 \to x_2$ and $x_3 \to x_4$.
The OPEs in \eqref{eq:phi-phi-OPE} then yield
\begin{align}
	\langle \Phi_a^\dagger(x_1) \Phi_b(x_2) &\Phi_c^\dagger(x_3) \Phi_d(x_4)\rangle_0 \sim \frac{\delta_{ab} \delta_{cd}}{|x_1-x_2|^{2 \Delta_\Phi}|x_3-x_4|^{2 \Delta_\Phi}} \nonumber\\
	&- (c_{\Phi \Phi}^M)^2 \frac{\mathcal{F}^{\mu \nu}_{ab} \mathcal{F}^{\rho \lambda}_{cd} \langle M_{\mu \nu}(x_2) M_{\rho \lambda}(x_4) \rangle_0}{|x_1-x_2|^{2 \Delta_\Phi - \Delta_M} |x_3-x_4|^{2 \Delta_\Phi - \Delta_M}}.
\end{align}
The first term is readily seen to yield a contribution of the $(\bm{h}^\ast \cdot \bm{h})^2$ in $F\mf$, similar to the $\langle \mathcal{S}_h^2\rangle_0^2$ term stemming from re-exponentiating the expanded $\eu^{-\mathcal{S}_h}$.

For the second term, we note that using the two-point function $\langle M_{\mu \nu}(x_2) M_{\rho \lambda}(x_4) \rangle_0 = \delta_{\mu \rho} \delta_{\nu \lambda} |x_2-x_4|^{-2 \Delta_M}$, we can perform the contraction $\mathcal{F}^{\mu \nu}_{ab} \mathcal{F}^{\rho \lambda}_{cd} \delta_{\mu \rho} \delta_{\nu \lambda} = \delta_{ac} \delta_{bd} - \delta_{ad}\delta_{bc}$.
Since the remaining integrals have integrands which are strictly positive functions, the second term gives a contribution of the form
\begin{equation} \label{eq:sh4-conn}
	\langle \mathcal{S}_h^4 \rangle_0 \sim - (c_{\Phi \Phi}^M)^2 C(L) \left[ (\bm{h} \cdot \bm{h}) (\bm{h}^\ast \cdot \bm{h}^\ast) - (\bm{h}^\ast \cdot \bm{h})^2 \right],
\end{equation}
where $C(L)> 0$ is an IR-divergent prefactor \ufps{which} depends on the cutoff length scale $L$ introduced earlier.
Importantly, this implies that the fourth-order contribution in $F\mf$ for $\bm{h}(\vec{x}) \equiv \bm{h} = \mathrm{const.}$ can be written to be of the form
\begin{equation}
	F\mf[h] \sim \dots + |h|^4 \left(D_1 + D_2 (\hat{\bm{h}}^\ast \cdot \hat{\bm{h}}^\ast)(\hat{\bm{h}} \cdot \hat{\bm{h}}) \right)+ \dots,
\end{equation}
where \eqref{eq:sh4-conn} implies that the prefactor $D_2>0$ (note that the sign of $D_1$ is left undetermined).

Importantly, the overall sign of the $|\hat{\bm{h}} \cdot \hat{\bm{h}}|^2$ term being positive implies that the free energy is minimized for configurations where the static components of the mean field satisfy $\hat{\bm{h}} \cdot \hat{\bm{h}} = 0$, i.e. they can be written as $\hat{\bm{h}} = (\hat{\bm{u}} + \iu \hat{\bm{v}})/\sqrt{2}$, with the two orthonormal vectors $\hat{\bm{u}},\hat{\bm{v}}$ such that $\hat{\bm{u}} \cdot \hat{\bm{u}} = \hat{\bm{v}} \cdot \hat{\bm{v}} = 1$ and $\hat{\bm{u}}\cdot\hat{\bm{v}} = 0$.
Considering that $\hat{\bm{h}}$ ultimately determines the vectorial nature of the monopole expectation value  $\langle \bm{\Phi} \rangle$ in the ordered phase (which can be identified with the Néel order parameter), we hence conclude that non-collinear intralayer spin order is energetically preferred [the generic form $\hat{\bvec{h}} = \hat{\bvec{u}} + \iu \hat{\bvec{v}}$ is seen to give rise to spin spiral ordering with basis vectors $\hat{\bm{u}},\hat{\bm{v}}$ using Eq.~\eqref{eq:spin_spin}].

\subsection{VBS monopoles}

The analyses for the VBS monopoles are in parallel.
We again perform a Fourier expansion on both sides of \eqref{eq:target}, yielding 
\be
\begin{split}
\sum_{\Q'}h_1^a(\Q')\eu^{\iu \Q'\cdot \x}= & \jv^2\ \fv^a(\x) \int d^3y\frac{e^{-i\q_a\cdot \vec{y}/2}}{|y-x|^{2\Delta_\Phi}} \\
& \times \sum_{\Q} h_1^a(\Q) \int d^3z\  \frac{\eu^{\iu \Q\cdot \vec{z}} }{|z-y|^{2\Delta_\Phi}}.
\end{split}
\label{eq:consistency3}
\ee
Taking into account the two possible singular contributions to the integrals at $\Q=\q_a/2$ and $\Q=0$, respectively, we get
\be
\begin{split}
&\ \sum_{\Q'}h_1^a(\Q')\eu^{\iu \Q'\cdot \x} \\
 =  &\ c \jv^2   \left|\frac{q}{2L}\right|^{2\Delta_\Phi-3}  \left[h_1^a(0)+h_1^a\left(\frac{\q_a}{2}\right)\eu^{\iu \q_a\cdot \x/2}\right] +\cdots,
\end{split}
\ee
For a nontrivial solution to exist, i.e., when $h_1^a(\x)\neq 0$, we must have $c\jv^2|q/2L|^{2\Delta_\Phi-3}=1$.
It immediately follows that $\jv$ exhibits the same scaling
\begin{equation} \label{eq:jv}
   \jv \sim (a_\mathrm{m} L)^{\Delta_\Phi - 3/2} 
\end{equation}
as $\js$ in \eqref{eq:js}.
Back to real space, the minimal description of the mean field is thus $
h_1^a(\x)= \tilde{h}_1^a(0)+\tilde{h}_1^a\left(\frac{q}{2}\right)\eu^{\iu \q_a\cdot \x/2}$
where $q=|q_a|$ is again independent of $a$. 
Similarly, the mean field for the other layer can be derived using \eqref{eq:coupled_h}:
\be
h_2^a(\x)\approx \tilde{h}_2^a(0) +\tilde{h}_2\left(-\frac{q}{2}\right)\eu^{-\iu \q_a\cdot \x/2}.
\ee
These four Fourier components in the two layers are related to each other through \eqref{eq:coupled_h}:
\be
\tilde{h}_1^a\left(\frac{q}{2}\right)=-A \tilde{h}_2^a(0),\quad \tilde{h}_2^a\left(-\frac{q}{2}\right)=-A \tilde{h}_1^a(0),
\ee
$A\gg 1$ is the same as defined in \eqref{eq:A}.   Combining with the mean field equations \eqref{eq:self-con-S}, we arrive at 
\be
\begin{split}
& \bk{{\Phi}_{1}^a(x)}\mf= \left( A\tilde{{h}}^a_1(0)-\tilde{{h}}^a_2(0) \eu^{\iu \q_a \cdot \x/2}\right)/\js,\\
& \bk{{\Phi}^a_{2}(x)}\mf= \left( A\tilde{{h}}_2^a(0)-\tilde{{h}}_1^a(0) \eu^{-\iu \q_a \cdot \x/2}\right)/\js.\\
\end{split}
\ee
Similar to the spin monopole case, we observe that the first terms in the two equations above dominate, and there are some corrections with modulations of moiré scale. The relation between $\tilde{h}_1^a(0)$ and $\tilde{h}_2^a(0)$ can also be obtained by going to quartic order perturbation, giving rise to $\tilde{{h}}_1^a(0)=s\tilde{{h}}_2^a(0)$ with $s$ a constant phase factor.
More details can be found in Appendix~\ref{app:spin}.  

We would like to comment that, within a microscopic theory, terms which gives rise to a coupling of the VBS order parameters involve four spin operators.
In contrast, the spin monopole tunneling can be generate from an interlayer spin-spin interaction.
In any setting where interlayer interactions are weak, we hence expect the interlayer tunneling of the spin monopoles to be dominant, and therefore instabilities with VBS order to be less likely.
{\section{Mean-field theory at strong coupling: Local density approximation} \label{sec:strong}}

In the limit of strong interlayer couplings $J_a \gg |\nabla f_a /f_a|^{2 \Delta_\Phi-3} \sim (1/a_\mathrm{m})^{3-2 \Delta_{\Phi}}$ (recall that we take $2 \Delta_\Phi < 3$ which amounts to assuming that interlayer monopole tunneling term is relevant), the spatial variations of the interlayer coupling due to twisting are on a much larger scale compared to the characteristic length scale set by the coupling strength $J_\alpha$.

\ufps{We note that the use of IR conformal field theory in the strong coupling limit is justified: this low-energy theory is expected to apply at energy scales $\Lambda \ll 1/a$ much smaller than the inverse lattice constant of the parent triangular lattice.
Recall that the moiré lattice constant $a_m \gg a$ and thus $ (1/a_m)^{3-2\Delta_\Phi} \ll (1/a)^{3-2\Delta_\Phi}$.
We conclude that these inequalities show that there exists a parameter regime such that the strong-coupling limit holds, but the typical energy scale $\Lambda \sim J_a^\frac{1}{3-2 \Delta}$ set by the interlayer coupling is small compared to the UV cutoff determined by the inverse parent lattice scale $1/a$, allowing us to use the low-energy conformal field theory to study the strong-coupling limit.
We also note that there is also \emph{another} parameter regime where $J_a \gg (1/a)^{3-2 \Delta_{\Phi}}$, implying that $J_a$ is the dominant scale in the problem and larger than the UV cutoff scale, necessitating the use of microscopic lattice models to study strong interlayer interactions.
However, this scenario seems less relevant to van der Waals heterostructures, where the interlayer interactions are generically weaker than intralayer interactions, both being small compared to the UV cutoff provided by the inverse lattice constant (we emphasize that the interlayer interactions being ``weak'' compared to intralayer interactions does not contradict the \emph{strong-coupling limit} defined earlier, where the interlayer interaction scale is large compared to moiré lattice modulations).}

\ufps{In this strong-coupling limit}, we may perform a \emph{local approximation} where we first obtain monopole expectation values as a function of the background fields from a \emph{local} scaling form of the compact \qed{} free energy in the presence of (large) background fields.
Within this local approximation, the self-consistency equations for distinct points in space-time decouple and can be solved straightforwardly.

To this end, we first note from $\mathcal{S}_h[\bm{h}_l]$ that the mean fields $\bm{h}_l$ have the scaling dimension $[h_l] = 3 - \Delta_\Phi$, such that a symmetry-allowed ansatz for the mean-field free energy $F\mf^\mathrm{loc}[\bvec{h}_l]$ (for each layer $l$) in the local approximation reads
\begin{equation}
    F\mf^\mathrm{loc}[\bvec{h}_l] = -\int \du^3 x \ \frac{2 c_\mathrm{loc}}{\chi} |\bvec{h}_l(x)|^\chi,
\end{equation}
where $c_\mathrm{loc}$ is some real constant and $\chi = 3/(3-\Delta_\Phi)$ from scaling arguments, with the form of the prefactor chosen for later convenience. 

Using that $\langle \Phi_{la} \rangle\mf = - \delta F\mf/\delta h_{l}^{a\ast} = c_\mathrm{loc} \hat{h}_{l}^a |\bvec h_l|^{\chi-1}$ (with $\hat{h}_{l}^a$ denoting the $a$-component of the complex unit vector $\hat{\bm{h}}_l = \bm{h}_l / |\bvec h_l|$), the self-consistency equations \eqref{eq:self-con-S} can be written as 
\begin{subequations}
\begin{align}
    h_1^a(x) &= - J_a f_a(x) c_\mathrm{loc} \hat{h}_2^a(x) |\bvec{h}_{2}(x)|^{\chi-1}  \\
    h_2^a(x) &= - J_a f_a^\ast(x) c_\mathrm{loc} \hat{h}_1^a(x) |\bvec{h}_{1}(x)|^{\chi-1} ,
\end{align}
\label{eq:self-con-sc}
\end{subequations}
with no implicit summation over $a$, and recall that $J_{1,2,3} \equiv \jv$, $J_{4,5,6} \equiv \js$, and $f_{1,2,3}(x) \equiv f_\mathrm{v}^{1,2,3}(x)$ and $f_{4,5,6} \equiv f_\mathrm{s}$.

\subsection{Spin monopoles}

We first focus on order in the spin sector (i.e. $\js \neq 0,\jv = 0$) such that $\bvec{h}$ in each layer can be replaced by a three-dimensional complex vector $(h^4,h^5,h^6)^\top$.
First focusing on magnitudes $|\bvec{h}_l|$ and substituting \eqref{eq:self-con-sc} into each other yields
\begin{equation} \label{eq:hl-sc}
    |\bvec{h}_l(x)| = |\js c_\mathrm{loc} f(x)|^\frac{3-\Delta_\Phi}{3-2 \Delta_\Phi}
\end{equation}
for $l=1,2$.
Using this in Eq.~\eqref{eq:self-con-sc}, we find that the \emph{relative} phase factor of the two unit vectors $\hat{\bvec{h}}_l$ is fixed by
\begin{equation} \label{eq:h1fromh2}
    \hat{\bm{h}}_1 = \exp\left[\iu \left(\pi + \arg \js + \varphi_\mathrm{s}(x) + \arg c_\mathrm{loc} \right)\right] \hat{\bm{h}}_2,
\end{equation}
where we have introduced $\varphi_\mathrm{s,v}(x) \equiv \arg f_\mathrm{s,v}(x)$.
Note that, considering the case of a homogenous bilayer (i.e. $f_\mathrm{s}(x) \equiv 1$) with a ferromagnetic interlayer coupling $\js < 0$, we expect the two Néel vectors (and so the mean fields $\bm{h}$) in the two layers to be aligned, which allows us to \emph{a posteriori} fix the sign of the constant $c_\mathrm{loc}>0$.
Above result is also readily understood when considering the expectation value of the interlayer tunneling term and rewriting using \eqref{eq:self-con-S},
\begin{equation}
\begin{split}
	\langle \mathcal{L}_1 \rangle\mf & \sim \js f(x) \langle\bm{\Phi}_1^\dagger\rangle\mf\cdot \langle\bm{\Phi}_2\rangle\mf \\
	& \sim \frac{1}{\js f(x)} |h(x)|^2 \hat{\bm{h}}_1 \cdot \hat{\bm{h}}_2^\ast + \hc,
\end{split}
\end{equation}
such that all phase fluctuations cancel out upon substituting \eqref{eq:h1fromh2}.

While we have thus fixed the magnitude of the mean fields (and thus of the magnetic order parameters, identified with the monopole operators) and their relative phase factor, we note that the self-consistency equations in the local approximation have a \emph{local} $\mathrm{U}(3)$ redundancy:
Given a configuration $\{\bm{h}_1(x),\bm{h}_2(x)\}$ which satisfies Eqs.~\eqref{eq:self-con-sc}, the configuration $\{G(x) \bm{h}_1(x),G(x) \bm{h}_2(x)\}$ with an arbitrary matrix field $G(x) \in \mathrm{U}(3)$ is a solution as well.
This local redundancy is unphysical as it does not correspond to a symmetry of the system and thus will be lifted by corrections to the purely local approximation made above.

In particular, we posit that there is an intrinsic stiffness to the system which energetically favors order parameter textures with small gradients.
Using $\SO(6)$-symmetry and the above-derived scaling of $\bm{h}_l$, a corresponding mean-field stiffness term in the free energy may be written as a sum over the two individual stiffness terms in each layer,
\begin{equation} \label{eq:gradients}
	F\mf^{\nabla^2}[\bm{h}_1,\bm{h}_2] = \sum_{l=1,2} \rho \int \du^3 x \ \left|\partial_\mu h_{l}^{a\ast}\ \partial^\mu h_{l}^a \right|^\gamma,
\end{equation}
where $\rho >0$ is some dimensionless constant and $\gamma= 3 /(8-2 \Delta_\Phi)$ from scaling (note summation over $a=1,\dots,6$ is implied).
Clearly, \eqref{eq:gradients} does not support the local $\mathrm{U}(3)$ redundancy found earlier, and is minimized \emph{in a single layer} through a uniform $\bm{h}(x) = \mathrm{const}.$ -- however, choosing \emph{both} $\bm{h}_{1}(x)$ and $\bm{h}_2(x)$ constant is no longer a self-consistent solution as is readily verified using Eqs.~\eqref{eq:self-con-sc}.
We here assume that the stiffness is parametrically small compared to the density term, $\rho \ll c_\mathrm{loc}$, such that the gradient terms can be treated as a perturbation which selects a favorable configuration out of the $\mathrm{U}(3)$-locally-degenerate manifold, without qualitatively altering the nature of the thus selected configuration.
Hence, we plug in the solution to the self-consistency equations, given by Eqs.~\eqref{eq:hl-sc} and \eqref{eq:h1fromh2} to rewrite the gradient term as a functional $F\mf^{\nabla^2}[\bm{h}_1,\bm{h}_2] \equiv F\mf^{\nabla^2}[\hat{\bvec{h}}_2]$ of the to-be-determined unit-vector-valued field $\hat{\bvec{h}}_2(x)$.
We find the resulting form difficult to extremize analytically due to the non-analycity of the gradient term.

Instead, we note that $\gamma = 3 /(8-2\Delta_\Phi) < 1$ and thus the modulus $|\dots|^\gamma$ is concave, such that Jensens inequality $|X|^\gamma + |Y|^\gamma \leq 2^{1-\gamma} |X+Y|^\gamma$ holds.
Hence, we resort to extremizing an \emph{upper bound} for $F\mf^{\nabla^2}$,
\begin{align}
    &\frac{F\mf^{\nabla^2}}{2^{1-\gamma} \rho} \leq  \int \du^3 x \Big[ \big| |h(x)|^2 \big( 2 \partial_\mu \hat{\bm{h}}_2^\ast \cdot \partial^\mu \hat{\bm{h}}_2 - \iu (\partial_\mu \varphi_\mathrm{s}) \big(\hat{\bm{h}}_2^\ast \cdot \partial^\mu \hat{\bm{h}}_2 \nonumber\\
    &-\hc\big) +\partial_\mu \varphi_\mathrm{s}\partial^\mu \varphi_\mathrm{s}\big) + 2 \partial_\mu |h| \partial^\mu |h| \big|^\chi + \lambda(x) \big( \hat{\bm{h}}^\ast_2 \cdot \hat{\bm{h}}_2 -1 \big)\Big] \label{eq:bound-grad}
\end{align}
where the short form $|h| \equiv |\bvec{h}_l|$ as in \eqref{eq:hl-sc},
and we use ``$\cdot$'' to denote the dot product of $\SO(3)_\mathrm{spin}$-vectors.
Note that $\hat{\bm{h}}_2^\ast \cdot \partial_\mu \hat{\bm{h}}_2 = - \hat{\bm{h}}_2 \cdot \partial_\mu \hat{\bm{h}}^\ast_2$ holds due to normalization, the latter being enforced by introducing a real Lagrange multiplier field $\lambda(x)$. Here we have chosen to write \eqref{eq:bound-grad} in a manifestly hermitian form.

Considering the right-hand side of \eqref{eq:bound-grad}, we note that due to the chain rule, it is sufficient to only consider the term in the square brackets when extremizing (by varying with respect to $\hat{\bm{h}}_2$, $\hat{\bm{h}}_2^\ast$ independently).
Using a redefined (real) Lagrange multiplier field $\tilde{\lambda}$, one thus obtains the differential equation
\begin{equation}
	0=-2 \partial_\mu \partial^\mu \hat{\bm{h}}_2 - 2 \iu \partial_\mu  \varphi_\mathrm{s}\partial^\mu \hat{\bm{h}}_2 - \iu \partial_\mu \partial^\mu \varphi_\mathrm{s}\hat{\bm{h}}_2 + \tilde{\lambda} \hat{\bm{h}}_2,
\end{equation}
and similarly for $\hat{\bm{h}}_2^\ast$.
Making the ansatz $\hat{\bm{h}}_2(x) = \hat{\bm{u}} \eu^{\iu g(x)}$ for some constant complex unit vector $\hat{u}$ such that the normalization constraint (retrieved by varying with respect to the Lagrange multiplier) is satisfied, the real function $g(x)$ is determined via
\begin{equation} \label{eq:eq-for-g}
	0 = - 2 \iu \partial_\mu \partial^\mu g + 2 \partial_\mu g \partial^\mu g + 2 \partial_\mu \varphi_\mathrm{s}\partial^\mu g - \iu \partial_\mu \partial^\mu \varphi_\mathrm{s}+ \tilde{\lambda}
\end{equation}
Adding and subtracting \eqref{eq:eq-for-g} and its complex conjugate, we find the partial differential equations
\begin{subequations}
\begin{align}
    0 &= 2 \partial_\mu g \partial^\mu g + 2 \partial \varphi_\mathrm{s} \partial^\mu g + \tilde{\lambda} \label{eq:PDE_lambda}, \\
    0 &= 2 \partial_\mu \partial^\mu g + \partial_\mu \partial^\mu \varphi_\mathrm{s}. \label{eq:PDE_g}
\end{align}
\end{subequations}
While the first equation can be solved to determine the Lagrange multiplier $\tilde{\lambda}$, the latter can be used to determine $g(x)$.
\emph{One particular solution} consists in choosing $g_\mathrm{p}(x) = - \varphi_\mathrm{s}(x) /2$ and $\tilde\lambda = - \partial_\mu \varphi_\mathrm{s}\partial^\mu \varphi_\mathrm{s}/ 2$ \footnote{One may worry that the choice of $p=-1/2$ implicitly depends on using the manifestly hermitian form of \eqref{eq:bound-grad}, however writing $\hat{h}_2^\ast \cdot \partial^\mu \hat{h}_2 - \hc \equiv (1+\alpha) \hat{h}_2^\ast \cdot \partial^\mu \hat{h}_2 - (1-\alpha) \hat{h}_2 \cdot \partial^\mu \hat{h}_2^\ast$ and repeating the calculation for a generic $\alpha$ shows that $p=-1/2$ independent of $\alpha$.}.

However, we note that $\varphi_\mathrm{s}(x) = \arg \fs (x)$ has branch cuts that connect pairs of zeros $x^\pm$ of $\fs$ at which $\fs(x^\pm) = 0$ and upon encircling $\fs$ has a positive (negative) winding number $\mathrm{Ind}_{\fs}(x^\pm) = \pm 1$. 
Note that here, we use conventions such that $\arg \fs \in [-\pi,\pi]$.
Here, we may take $x^+ = (1 / \sqrt{3},0) a_\mathrm{m}$ and $x^- = (2 / \sqrt{3},0) a_{\mathrm{m}}$, as marked by a red cross in Fig.~\ref{fig:f}, and by spatial periodicity of $\fs(x)$, all further zeros of positive (negative) winding number are obtained through translations by lattice vectors of the magnetic Brillouin zone 

While the branch cut above leaves $\eu^{\iu \varphi_s(x)}$ single-valued, the function $g_\mathrm{p}(x) = - \varphi_\mathrm{s}(x)/2$ features a discontinuity across these branch cuts and hence the $\eu^{\iu g_\mathrm{p}(x)}$ is multivalued at these branch cuts (i.e. there is a jump $\pi/2 \to -\pi/2$ when encircling a zero of $\fs$ with positive chirality, $x^+$).
Since the phase of $\hat{\bvec{h}}_l(x) = \hat{\bvec{u}} \eu^{\pm \iu g(x)}$ determines the phase of $\langle \bvec{\Phi}_l \rangle$, which in turn determines the local orientation of the ordered spins via Eq.~\eqref{eq:spin_spin}.
This is an \emph{unphysical} discontinuity, and hence the choice of $g_\mathrm{p}(x) = - \varphi_\mathrm{s}(x)/2$ is not admissible.
To remedy this, we recall that above choice corresponds to a \emph{particular} solution, and in fact any $g(x) = g_\mathrm{h}(x) + g_\mathrm{p}(x)$ is a solution to \eqref{eq:PDE_g}, where $g_\mathrm{h}$ is a solution to Laplace's equation, $\partial_\mu \partial^\mu g_\mathrm{h}(x) = 0$.
A boundary condition for $g_\mathrm{h}$ consists in demanding that it has the same spatial symmetries as $g_\mathrm{p}(x)$.

One can make progress by noting that $\arg(x,y) \equiv \arg(x+\iu y) = \Im \log(x+\iu y)$ is a solution to Laplace's equation and features a branch cut on the negative half of the $x$-axis (note that other conventions may be chosen), such that $\lim_{y\to \pm 0} \arg(x,y) = \pm \pi$ for $x < 0$. 
Hence, by superposing two such solutions to define a vortex-pair function
\begin{equation}
    G_\mathrm{vp}^{(3)}(x,y) \equiv \arg \left(-x+\frac{a_\mathrm{m}}{\sqrt{3}},y\right) - \arg\left(-x+ \frac{2 a_\mathrm{m}}{\sqrt{3}},y\right),
\end{equation}
one obtains a branch cut between the points $(1/\sqrt{3},0) a_\mathrm{m}$ and $(2 / \sqrt{3},0) a_\mathrm{m}$ where $G_\mathrm{vp}^{(3)}(x,y)$ winds by $-2 \pi$ (clockwise direction) around the former point and by $+ 2\pi$ around the latter.
One may form similar functions $G_\mathrm{vp}^{(1)}(x,y)$ and $G_\mathrm{vp}^{(2)}(x,y)$ with branch cuts connecting the singular points $a_\mathrm{m} (1/(2 \sqrt{3}),1/2)$ and $a_\mathrm{m} (1/ \sqrt{3},1)$ [$a_\mathrm{m} (-1/(2 \sqrt{3}),1/2)$ and $a_\mathrm{m} (- 1/ \sqrt{3},1)$, respectively].

We may then take the homogeneous solution as a linear superposition of the vortex-pair functions translated by the lattice vectors of the magnetic moiré Brillouin as given in \eqref{eq:latvec_mmbz},
\begin{equation}
    g_\mathrm{h}(\vec x) = \frac{1}{2} \sum_{m,n \in \mathbb{Z}} q_{mn}^{(\alpha)}  G^{(\alpha)}_\mathrm{vp}(\vec x + n \vec b_1^{(\mathrm{m})} + m \vec b_2^{(\mathrm{m})}),
\end{equation}
where the coefficients $q_{mn}^{(\alpha)} =\pm 1$ and should not be confused with the moiré reciprocal lattice vectors $\vec q_i$. 
The prefactor of 1/2 is chosen such that at the discontinuity  $g_\mathrm{h}(\vec x)$ changes $\pm \pi /2 \to \mp \pi/2$.

Considering the full function $g(x) = g_\mathrm{h}(x) + g_\mathrm{p}(x)$, one thus finds that by choosing the value of $q_{mn}^{(\alpha)} = \pm 1$,  one may remove the half branch cut due to the $-\varphi_\mathrm{s}/2$ in one layer (such that the phase of $\eu^{\iu g(x)}$ does not wind around a vortex pair) and obtain a full branch cut (associated with $\pm 2 \pi$ winding around two paired zeros) in the other layer such that $\eu^{\iu \varphi_\mathrm{s}(x) + g(x)}$ is single-valued and thus physical.
Note that reversing the sign of $q_{mn}^{(\alpha)}$ then corresponds to placing the (physical) vortex pair in the opposite layer. 
Within this approach, each choice of $q_{mn}^{(\alpha)}$ is an allowed solution, i.e. for each pair of zeros of $\arg f_\mathrm{s}(x)$ we have the choice of placing a vortex pair in the upper or lower layer. 
While these are all \emph{local} extrema of the free energy (as they solve Eqs.~\eqref{eq:PDE_lambda} and \eqref{eq:PDE_g}), the total energy of these configurations may differ.

Considering \eqref{eq:gradients}, it is reasonable to expect that the energetically favored configurations consist in placing all vortex pairs in either the top or bottom layer such that the gradient term for the other layer is minimized (i.e. $q_{mn}^{(\alpha)} \equiv +1$ or $q_{mn}^{(\alpha)}\equiv-1$ for all $m,n,\alpha$).
Consequently, we expect the interlayer exchange symmetry to be spontaneously broken in this scenario.

To summarize this section, when the tunnelings of spin monopoles are present, in this strong-coupling, local approximation, we have found a magnetic vortex lattice of tunable moiré scale.
The expectations values of the spin monopoles in the two layers differ by a spatially modulating phase $\bk{\bm{\Phi}_2}\mf/\bk{\bm{\Phi}_1}\mf=\eu^{\iu \arg \fs}$, which winds by $\pm 2\pi$ around the moiré downward/upward triangles, as depicted in Fig.~\ref{fig:vortex_lattice}.


\subsection{VBS monopoles}

In the VBS sector ($\jv \neq 0$), we can proceed similarly to the spin sector.
Now, we consider the first three components of the (vectorial) mean fields $\bm{h}_1,\ \bm{h}_2$.
Recalling that all $\fv^a$'s are different, we take the magnitude of each component of $h_l^a$ in Eqs.~\eqref{eq:self-con-sc} and \eqref{eq:self-con-S}. Using that $|f_a(x)| = 1$ we hence find that the magnitude of the mean fields is constant,
\begin{equation} \label{eq:hl-abs-vbs}
	|h_{l}^a(x)| = | \jv c_\mathrm{loc}|^\frac{3 - \Delta_\Phi}{3- 2 \Delta_\Phi}
\end{equation}
for $l=1,2$ and $a=1,2,3$, and the constant $c_\mathrm{loc} > 0$ as argued earlier.
Performing a polar decomposition of each mean field component, $h_{l}^a = |h_{l}^a| \eu^{\iu \vartheta_{l}^a}$, we find that the phases in the two layers must satisfy
\begin{equation}
	\vartheta_1^a(x) = \pi + \arg \jv + \arg c_\mathrm{loc} + \varphi^a(x) +\vartheta_{2}^a(x),
\end{equation}
where $\varphi^a(x) \equiv \arg f^a_\mathrm{v}(x) = \vec{k}_a \cdot \vec{x}$.
Similar to the case of spin monopoles discussed in the previous section, the  self-consistency equations admit a \emph{local} redundancy in the VBS sector.
However, due to flavor-dependence of the interlayer monopole tunneling function, the space of locally degenerate solutions is smaller compared to spin case: 
In the VBS sector, the self-consistency equations (and their solutions) remain invariant under three independent $\Uone$ rotations $h_{l}^a \mapsto \eu^{\iu \xi^a} h_{l}^a$ with some real field $\xi^a(x)$.
This unphysical local redundancy is again lifted in first order by a gradient-type term as given in \eqref{eq:gradients}.
We again consider an upper bound for the gradient term using Jensen's inequality (also taking $h^{4,5,6} = 0$).
Inserting $h_{l}^a = |h_{l}^a| \eu^{\iu \vartheta_l^a}$ the problem reduces to minimizing the right-hand side of
\begin{align}
	\frac{F\mf^{\nabla^2}}{2^{1-\gamma}\rho} \leq \int \du^3 x \Big|\sum_{a=1,2,3} |h|^2 \big( &2 \partial_\mu \vartheta_2^a \partial^\mu \vartheta_2^a + 2 \partial_\mu \varphi^a \partial^\mu \vartheta_2^a\nonumber\\
	&+ \partial_\mu \varphi^a \partial^\mu \varphi^a \big) \Big|^\gamma \label{eq:bound-grad-vbs}
\end{align}
with respect to $\vartheta_2^a$, and $|h|=|h_{l}^a(x)|$ given in \eqref{eq:hl-abs-vbs}.
Note that no Lagrange multiplier is required as the polar decomposition ensures normalization of $h_{l}^a/|h_{l}^a|$.
Again, by the chain rule, it is sufficient to consider the expression in parenthesis.
Varying with respect to $\vartheta_2^a$, one obtains
\begin{equation} \label{eq:pde_vbs}
	0 = -2 \partial_\mu \partial^\mu \varphi_a(x) - 4 \partial_\mu \partial^\mu \vartheta^a_2(x).
\end{equation}
Note that since $\partial_\mu \partial^\mu \varphi_a(x) = \partial_\mu \partial^\mu (\vec k_a \cdot \vec x) \equiv 0$, Eq.~\eqref{eq:pde_vbs} reduces to Laplace's equation $\partial^\mu \partial_\mu \vartheta_2^a(x) = 0$.
Aiming to find a \emph{global} minimum of the RHS in \eqref{eq:bound-grad-vbs}, we find that choosing $\vartheta^a_2(x) = -\varphi_a(x)/2$ gives a lower bound than the constant solution $\theta^a_2 = \mathrm{const.}$

We stress that, in contrast to the case of spin monopoles discussed in the previous subsection, the phase $\varphi^a(x) \equiv \vec k_a \cdot \vec x$  does not lead to zeros in $\eu^{\pm \iu \varphi^a(x)/2}$, but rather doubles the wavelength of the spatial periodicity of the modulation of the phase of the VBS order parameter.

Note that there is a freedom to pick a global $\Uone$ phase $\phi_0^a$ per flavor (analogous to the choice of $\hat{u}$ in the spin case).
The mean fields can hence be written as
\begin{subequations}
	\begin{align}
	h_{1}^a(x) &= |\jv c_\mathrm{loc}|^\frac{3 - \Delta_\Phi }{3-2 \Delta_\Phi} \eu^{\iu (\pi + \arg \jv + \varphi_a(x)/2 + \phi_0^a)} \\
	h_{2}^a(x) &= |\jv c_\mathrm{loc}|^\frac{3 - \Delta_\Phi }{3-2 \Delta_\Phi} \eu^{\iu (-\varphi_a(x)/2 + \phi_0^a)}.
\end{align}
\end{subequations}
Consequently, the VBS order parameters, obtained from inserting above result in \eqref{eq:self-con-VBS}, remain constant in magnitude throughout the moiré lattice, but feature oscillating phases with wavevectors $\vec k_a = \ufps{-}\vec K_a / 2$ corresponding to half of the moiré lattice's reciprocal lattice vectors. This further leads to modulating phase difference between the monopole expectation values in the two layers $\langle \Phi_{1a}\rangle\mf / \langle \Phi_{2a}\rangle\mf =\exp[\iu(-\pi -\arg\jv+\varphi_a(x))]$. Namely, \zx{on the one hand,} in certain regions where the phase difference is small, the singlets in the two layers almost lie on top of each other; on the other hand, where the phase difference is near $\pi$, the singlets in the two layers avoid each other.

\section{Discussion and Conclusion}
\label{sec:discussion}

We conclude the paper with a summary of our results, a  discussion pertinent to underlying assumptions of our study, and an outlook.

\subsection{Summary}

\begin{figure}
    \centering
    \includegraphics[width=\columnwidth]{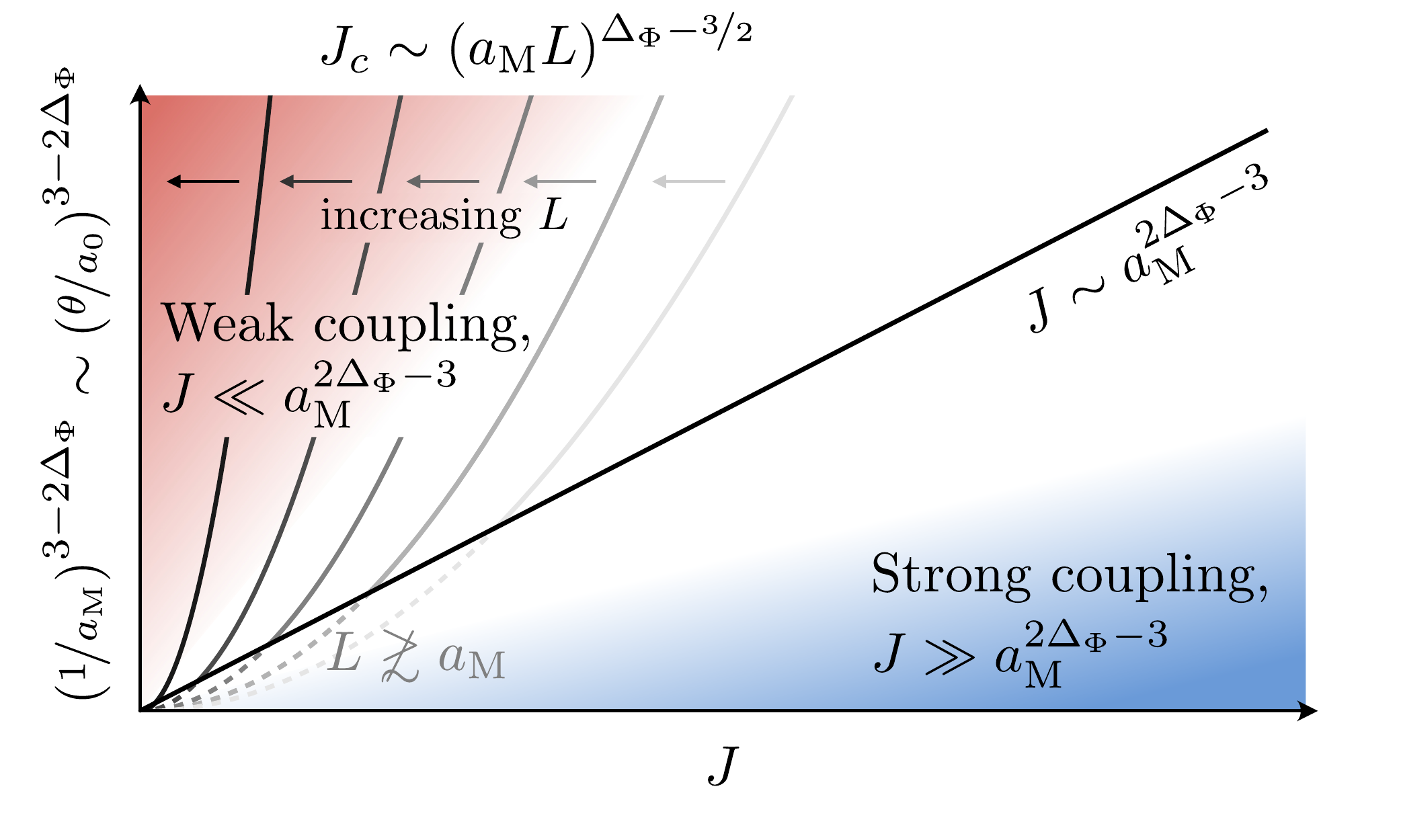}
    \caption{\label{fig:scaling_overview}Illustration of the relation between the weak coupling (red shading) and the strong coupling limits (blue shading), and scaling of the critical $J_c$, as a function of $(1/a_\mathrm{m})^{3-2\Delta_\Phi}$ and $J \equiv J_\alpha$. Note that increasing values on the ordinate corresponds to \emph{decreasing} the size of the moiré unit cell. The dashed lines indicate the parameter regime for which the finite-size weak-coupling instability at $J_c \sim (a_\mathrm{m} L)^{\Delta_\Phi-3/2}$ implies that $L \ll a_\mathrm{m}$, which is deemed unphysical and vanishes in the thermodynamic $L\to \infty$ limit. Here, we use $a_0$ to denote the bare lattice constant of single-layer  triangular lattice.}
\end{figure}

In this work, we have shown that in bilayer systems of $\Uone$ Dirac spin liquids, interlayer monopole-monopole and mass-mass interactions are present in the low-energy field theory.
These terms constitute as \emph{relevant} perturbations to the strongly coupled conformal fixed point described by two copies of \qed{}.
{Employing a perturbative calculation informed by the low-energy CFT data, we find that an instability due to interlayer interaction persists in the twisted system, but its relevance is effectively reduced (i.e. $J_c \sim (a_\mathrm{m} L)^{\Delta_\Phi - 3/2}$, compared with $J_c \sim L^{-( 3- 2\Delta_\Phi)}$ in the homogeneous case).}

Motivated by the identification of monopole operators with antiferromagnetic Néel and VBS order parameters on the triangular lattice, we have focused on monopole-monopole interactions (likely the most relevant interlayer term) and developed a conformal mean-field theory framework which allows us to study ordered phases described by the proliferation of corresponding monopoles.
{Solving the mean-field equations at weak coupling, we recover the modified critical scaling for finite twist angles, and find that the induced spin/VBS order are mostly uniform, with small twist-induced modulations. We expect this weak-coupling approach} to be justified when the interlayer coupling is weak compared to the moiré modulations due to the twisting (red regime in Fig.~\ref{fig:scaling_overview}.
In the limit of strong interlayer couplings (or, equivalently, \emph{large} moiré unit cells achieved at small twist angles), $J\gg a_\mathrm{m}^{2 \Delta_\Phi -3}$, we employ a local approximation to show that the magnetic order parameter forms a moiré vortex lattice.

A qualitative overview of the two limits and resulting scaling laws is presented in Fig.~\ref{fig:scaling_overview}.
We remark that the mean-field solutions in the two regimes cannot be continuously connected to each other: in the weak-coupling case, the solutions \eqref{eq:summary_weak} exhibit a $\mathbb{Z}_2$ symmetry of $\bm{\Phi}_1\leftrightarrow -\bm{\Phi}_2^\dagger$, while at strong coupling, this symmetry is broken.
This suggests the presence of an additional transition, or even intervening phases, in the intermediate regime, constituting an interesting (albeit challenging) task for further study.

\subsection{Other studies of 
twisted spin liquids}
\label{sec:related}

There have been a few prior studies of twisted spin liquids.  Ref.~\onlinecite{Magic_cont} studied the twisted bilayer of the staggered flux state on the square lattice, which is also a candidate mean field state of a $\Uone$ Dirac  spin liquid, but the effect of monopoles were not taken into account.
In a similar vein, Ref.~\onlinecite{chenlado21} explored mean-field spinon band structures of twisted van der Waals magnets hosting $\Uone$ Dirac quantum spin liquids, finding a twist-induced gap opening and arguing that resulting band structures can be tuned upon magnetic encapsulation and applied magnetic fields.
Moving away from $\Uone$ Dirac spin liquids, we further mention that in Refs.~\onlinecite{mayhugh20} and \onlinecite{haskell} twisted versions of the  bilayer Kitaev honeycomb $\Ztwo$ spin liquid \cite{sei18} were studied.
While all these prior works are interesting, they differ fundamentally from our results as in our case we account fully for the nature of the Dirac spin liquid as a non-trivial conformal field theory.

\subsection{Conformal renormalization group flow for homogenous interlayer couplings}
\label{sec:conformal-rg}

In the study at hand, we have primarily focused on instabilities towards ordered phases due to a (simultaneous) proliferation of monopoles in both layers driven by the monopole interlayer tunneling terms.
However, as discussed in \ref{subsec:mass}, interlayer mass-mass couplings are also symmetry-allowed and thought to be relevant.
A key question thus pertains if (additional) instabilities can occur due to these additional couplings.
To this end, we have performed a perturbative renormalization group calculation to quadratic order, utilizing the operator product expansions admitted by the conformal nature of the \qed{} fixed point \cite{cardy96}.
Here, we consider the case of $\q \equiv 0$, i.e. homogenous interlayer couplings which occurs for trivial (AA) stacking, such that %
combining equations \eqref{eq:setup} and \eqref{eq:mass_setup} leads to the full Lagrangian, given by 
\begin{subequations}
\begin{align} \label{eq:L_aa_full}
\mathcal{L}= &\ \mathcal{L}_{\text{QED}_3}^{(1)}+\mathcal{L}_{\text{QED}_3}^{(2)} + \mathcal{L}^{(12)}_\mathrm{int}, \\
\mathcal{L}^{(12)}_\mathrm{int} = &\sum_{a=1}^3 \jv (\Phi_{1,a}^\dagger \Phi_{2,a}+\hc) +\sum_{a=4}^6 \js (\Phi_{1,a}^\dagger \Phi_{2,a}+\hc)\nonumber\\
& + g_1\sum_{i=1}^3 M_{1,i0}M_{2,i0}+g_2\sum_{j=1}^3 M_{1,0j}M_{2,0j}\nonumber\\
& + g_3 \sum_{i,j=1}^3 M_{1,ij}M_{2,ij}.
\end{align}
\end{subequations}
As mentioned in \ref{sec:review}, the low-energy theory $\mathcal{S} = \int \du^3 x \mathcal{L}$ is naturally endowed with a UV cutoff (the lattice spacing $a_0$ which bounds the separation of two (low-energy) operator insertions from below.
The conformal renormalization group is most naturally formulated in real space and proceeds by changing the the UV cutoff $a_0 \to b a_0$ with $b = (1+ \delta l) > 1$ (with $\delta l$ being infinitesimal) and subsequently integrating out operator insertion pairs (at coordinates $x$,$y$) with relative separations in the spherical shell defined by $b a_0 \geq |x-y| \geq a_0$.
We relegate a more detailed discussion to Appendix \ref{app:conformal-rg} and give the RG flow equations in terms of the dimensionless variables $\tilde{J}_{\mathrm{s},\mathrm{v}} = J_{\mathrm{s},\mathrm{v}} a_0 ^{3-2\Delta_\Phi}$ and $\tilde{g}_i = g_i a_0^{3-2 \Delta_M} $. 

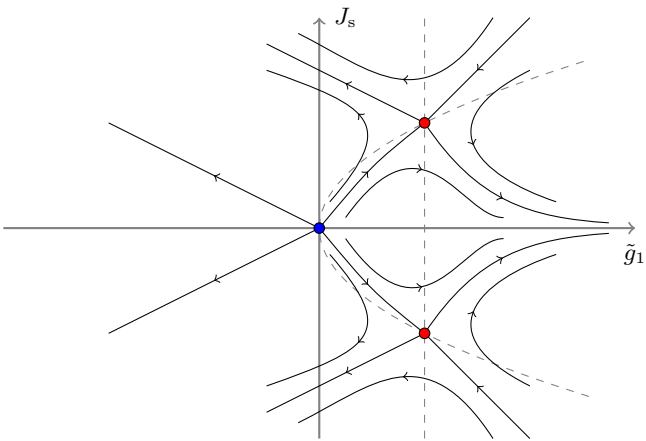
\begin{figure}[htbp]
    \centering
\begin{tikzpicture}[scale=0.7]
\draw[thick,->,gray] (-6,0)--(6,0);
\draw[thick,->,gray] (0,-4)--(0,4);
\draw[dashed,gray]   plot[smooth,domain=-3.2:3.2]
({\x*\x/2},\x);
\draw[dashed,gray] (2,-4)--(2,4);
\node at (6,-0.5) {$\tilde{g}_1$};
\node at (0.5,4) {$J_\mathrm{s}$};
\draw[->-=0.5] (0,0) .. controls (1,1.2) .. (2,2);
\draw[->-=0.5] (0,0) .. controls (1,-1.2) .. (2,-2);
\draw[->-=0.5] (4,4) .. controls (3,3) .. (2,2);
\draw[->-=0.5] (4,-4) .. controls (3,-3) .. (2,-2);
\draw[->-=0.5] (0,0) -- (-4,2);
\draw[->-=0.5] (0,0) -- (-4,-2);
\draw[->-=0.5] (2,2) -- (-1,3.5);
\draw[->-=0.5] (2,-2) -- (-1,-3.5);
\draw[->-=0.5] (2,2) .. controls (3,0.5) and (4,0.2).. (5.5,0.1);
\draw[->-=0.5] (2,-2) .. controls (3,-0.5) and (4,-0.2).. (5.5,-0.1);
\draw[->-=0.5] (0.5,0.2) .. controls (2,2.3) and (2.7,0.2).. (3.5,0.2);
\draw[->-=0.5] (0.5,-0.2) .. controls (2,-2.3) and (2.7,-0.2).. (3.5,-0.2);
\draw[->-=0.5] (4,3) .. controls (2,2) and (3,1).. (4.5,0.5);
\draw[->-=0.5] (4,-3) .. controls (2,-2) and (3,-1).. (4.5,-0.5);
\draw[->-=0.5] (3.3,4) .. controls (2,2) and (1,3).. (-0.4,3.7);
\draw[->-=0.5] (3.3,-4) .. controls (2,-2) and (1,-3).. (-0.4,-3.7);
\draw[->-=0.5] (0.2,0.5) .. controls (1.5,2) and (1,2.3).. (-1,3);
\draw[->-=0.5] (0.2,-0.5) .. controls (1.5,-2) and (1,-2.3).. (-1,-3);
\filldraw[fill=blue] (0,0) circle [radius=0.1];
\filldraw[fill=red] (2,2) circle [radius=0.1];
\filldraw[fill=red] (2,-2) circle [radius=0.1];
\end{tikzpicture}
    \caption{RG diagram for $\tilde{J}_\mathrm{v}=0=\tilde{g}_2=\tilde{g}_3$. See main text for discussions.}
    \label{fig:RG}
\end{figure}

We thus find the perturbative RG equations (up to quadratic order in $\tilde{J}_\mathrm{s,v}$ and $\tilde{g}_{1,2,3}$) to be of the form
\be
\begin{split} \label{eq:flow-eqns}
& \frac{\du \tilde{J}_\mathrm{s}}{\du l}=(3-2\Delta_{\Phi}) \tilde{J}_\mathrm{s} - c_1 (2 \tilde{J}_\mathrm{s} \tilde{g}_1 + 3 \tilde{J}_\mathrm{s} \tilde{g}_3),\\
& \frac{\du\tilde{J}_\mathrm{v}}{dl}=(3-2\Delta_{\Phi}) \tilde{J}_\mathrm{v} - c_1 (2 \tilde{J}_\mathrm{s} \tilde{g}_2 + 3 \tilde{J}_\mathrm{s} \tilde{g}_3),\\
& \frac{\du \tilde{g}_1}{\du l}=(3-2\Delta_{M}) \tilde{g}_1-2c_2 |\tilde{J}_\mathrm{s}|^2,\\
& \frac{\du \tilde{g}_2}{\du l}=(3-2\Delta_{M}) \tilde{g}_2-2c_2 |\tilde{J}_\mathrm{v}|^2,\\
& \frac{\du \tilde{g}_3}{\du l}=(3-2\Delta_{M}) \tilde{g}_3-c_2 (\tilde{J}_\mathrm{s} \tilde{J}_\mathrm{v}^* + \tilde{J}_\mathrm{v} \tilde{J}_\mathrm{s}^*).\\
\end{split}
\ee
Here, the constants $c_1 = \pi (c^\Phi_{\Phi M})^2 > 0$ and $c_2 = 4 \pi (c^M_{\Phi \Phi})^2>0$ are given in terms of the OPE coefficients as defined in \eqref{eq:phi-phi-OPE} and \eqref{eq:phi-M-OPE} and are left undetermined at this point. 
We find several nontrivial fixed points to the RG equations,
\be
\begin{split}
(\mathrm{i})\ & \tilde{J}_\mathrm{s}=\pm \frac{1}{\sqrt{4\tilde{c}_1\tilde{c}_2}},\ \tilde{J}_\mathrm{v}=0,\  \tilde{g}_1=\frac{1}{2\tilde{c}_1},\ \tilde{g}_2=\tilde{g}_3=0,\\
(\mathrm{ii})\ & \tilde{J}_\mathrm{v}=\pm \frac{1}{\sqrt{4\tilde{c}_1\tilde{c}_2}},\ \tilde{J}_\mathrm{s}=0,\  \tilde{g}_2=\frac{1}{2\tilde{c}_1},\ \tilde{g}_1=\tilde{g}_3=0,\\
(\mathrm{iii})\ & \tilde{J}_\mathrm{s}=\tilde{J}_\mathrm{v}=\pm \frac{1}{\sqrt{10\tilde{c}_1\tilde{c}_2}},\   \tilde{g}_1=\tilde{g}_2=\frac{1}{5\tilde{c}_1}=\tilde{g}_3,\\
(\mathrm{iv})\ & \tilde{J}_\mathrm{s} =-\tilde{J}_\mathrm{v}=\pm \frac{1}{\sqrt{10\tilde{c}_1\tilde{c}_2}},\   \tilde{g}_1=\tilde{g}_2=\frac{1}{5\tilde{c}_1}=-\tilde{g}_3,\\
\end{split}
\label{eq:FP}
\ee
where we have defined $\tilde{c}_1=c_1/(3-2 \Delta_\Phi)$ and $\tilde{c}_2=c_2/(3-2 \Delta_{M})$. 
Motivated by the expectation that on a microscopic (lattice) level, spin-spin interactions lead to bare interlayer couplings in the spin sector which are dominant compared to those in VBS sectors, we consider the example of $\tilde{J}_v=0=\tilde{g}_2=\tilde{g}_3$. 
A schematic resulting RG diagram is shown in Fig.~\ref{fig:RG}. 
The blue dot corresponds to the \emph{unstable} fixed point of two decoupled copies of QED$_3$, while the two red dots indicate the two \emph{critical} fixed points (i) as given in \eqref{eq:FP}. 
We observe that there are four distinct regimes of strong coupling, corresponding to (a1) $g_1\rightarrow -\infty$, $\js\rightarrow -\infty$; (a2) $g_1\rightarrow -\infty$, $\js\rightarrow \infty$; (b1) $g_1\rightarrow \infty$, $\js\rightarrow 0^-$;  (b2) $g_1\rightarrow \infty$, $\js\rightarrow 0^+$.


In the cases (b1) and (b2), the interlayer mass-mass couplings flow to strong coupling. 
In this regime, it can therefore be expected that the two masses are spontaneously generated (this simultaneous mass condensation in each layer would correspond to the fermion adjoint operators in each layer acquiring a finite expectation value, $\langle M_{1,i0} \rangle,\langle M_{2,i0} \rangle  \neq 0$), with $\tilde{g}_1 >0$ indicating that the signs of the generated masses are opposite.
As discussed in Ref.~\onlinecite{song19} using a Gross-Neveu type model, in a single layer, the generation of a spontaneous mass opens up a channel for a corresponding monopole to be proliferated, thus constituting a mechanism for a transition from the DSL to ordered phases.
In particular, the mass term $\pm M_{c0}$ will proliferate different $\Phi_{a}\pm i\Phi_b$ monopoles, where $(a-3,b-3,c)$ form an even permutation of $(1,2,3)$, and $\Phi_{a}\pm i\Phi_b$ implies non-collinear, but coplanar magnetic order. 
Note that while the perturbative RG flow with respect to the decoupled DSL fixed point appears to indicate that $\js \to 0^\pm$ in (b1) and (b2), it is to be expected that $\js$ (and potentially further monopole-monopole interactions) become relevant at the strong-coupling fixed point (i.e. they are \emph{dangerously irrelevant} couplings) and are crucial to the description of the confinement transition in the strong-coupling regimes (b1) and (b2) without referencing a Gross-Neveu theory as in Ref.~\onlinecite{song19}.

For the cases (a1) and (a2), a similar scenario as above (spontaneous mass generation which leads to the proliferation of the two monopoles) is conceivable.
However, the fact that $\js \to \pm \infty$ could also be taken to suggest that the physics in this regime is dominantly controlled by the monopole tunneling $\js$ growing strong, implying a \emph{direct} instability due to monopole tunneling.
As to the nature of the phase that is induced, two scenarios appear likely:

On the one hand, the interlayer monopole-monopole interaction can induce simultaneous monopole proliferation in both layers, thus realizing a (confined) long-range ordered phase.
This is the underlying assumption of our work at hand, justifying the use of mean-field theory (see Sec.~\ref{sec:mf}) in which the monopole operators acquire finite expectation values $\langle \bvec{\Phi}_1 \rangle,\langle \bvec{\Phi}_2 \rangle \neq 0$ which are in one-to-one correspondence with the Néel/VBS order parameters of the thus obtained ordered phases.

On the other hand, we note that the monopole-monopole tunneling term still preserves the total $\Uone_\mathrm{top}$ magnetic flux of the system.
Thinking about parton constructions (or the low-energy field theory \eqref{eq:qed3}), this could be taken to suggest that excitations coupled to the \emph{relative} gauge field $a_-^\mu \equiv a_1^\mu - a_2^\mu$ between the two layers (the flux of which is not conserved by the monopole tunneling term) become confined (the associated flux is no longer conserved), but the \emph{total} gauge field $a_+^\mu \equiv a_1^\mu + a_2^\mu$ remains in a deconfined state as the total $\Uone_\mathrm{top}$ flux conservation prohibits monopole proliferation.
The resulting state, obtained after such a ``partial confinement'' transition, would still support deconfined excitations.
However, as of now, it is unclear if (1) such a phase, which one might refer to as a ``bilayer spin liquid'', is \emph{stable} (single-fermion interlayer tunneling may lead to a gap opening in the spinon dispersion, and the resulting pure deconfined $\Uone$ gauge theory is unstable by Polyakov's mechanism \cite{poly75}) \emph{and} if (2) this phase is \emph{energetically competitive} compared with conventional (fully confined) ordered states.
A study of this intriguing scenario, necessary and plausible intermediate transitions involved, and the nature of resulting phases is left for future study.

The fixed points (2), (3) and (4) in Eq.~\eqref{eq:FP} can be analyzed analogously with identical RG diagrams looking the same upon appropriate reparametrizations.
Finally, a conformal RG study involving moiré modulations in the interlayer couplings is more involved and will be left for future research. 

\subsection{Applications and Outlook}

Potential material realizations include the van der Waals material 1T-TaS$_2$, which has been proposed to be either a Dirac or a $\mathbb{Z}_2$ spin liquid \cite{Law6996}.

We remark that generally in mean-field treatments, the tendency to order might be over-estimated.  However, experience with the analogous calculations in one lower dimension has shown them to be often very accurate. 
In the future, a study of fluctuations above mean-field level is nevertheless important to consider, though it is beyond the scope of this paper.
We do want to mention one possible way to study new disordered phases: Consider the couplings of bilayer masses as discussed in \ref{subsec:mass}, take an appropriate mean field ansatz such that the moiré low-energy spinon bands are flat (similar to what was found in the twisted bilayer staggered flux state on square lattice \cite{Magic_cont}).
The high density of low-energy states might help to suppress the monopoles and favor a spin liquid phase again.
As alluded in the previous subsection, one can conceive spin liquid phases that are intrinsic to the bilayer system, but it is unclear how stable those might be and if they can be energetically favored.

\ufps{Further, we note that the CFT as a low-energy theory for the coupled $\Uone$ Dirac spin liquids is applicable in the regime of small twist angles and thus large moiré lattice constants $a_\mathrm{m} \gg a$. If the twist angle is large (in the most extreme case $\theta = \pi/6$, since a twist by $0 \leq \theta \leq 2 \pi /6$ is equivalent to a twist by $2 \pi / 6 - \theta$ due to the $C_6$ lattice symmetry), there is no clear separation of scales $a_\mathrm{m} \sim a$ which would justify the use of the low-energy (continuum) theory and our results become uncontrolled.
Instead, microscopic (lattice) details will become important, an analysis of which we leave for further study.}

On the methodological side, we have established explicit operator-product expansions for the low-energy QED$_3$ theory of the Dirac spin liquid on the triangular lattice (similar expansions are readily obtained for e.g. the Kagome lattice), and developed a mean-field framework exploiting the conformal symmetry of the $\SU(4)$ DSL. Offering up a novel avenue for analytical treatments beyond large-$N$ calculations, our framework can be applied to the study of a wide array of perturbations to the Dirac spin liquid.

\acknowledgements
We thank Kasra Hejazi for collaborations on previous related works, Cenke Xu for helpful discussions, and Xue-Yang Song for correspondence on earlier work.
We further gratefully acknowledge discussions with {Max Metlitski} as well as Yin-Chen He and Chong Wang (ZXL), David Simmons-Duffin and Grant Remmen (UFPS). This work is supported in part by the Simons
Collaborations on Ultra-Quantum Matter, grant 651440 (ZXL) from the Simons
Foundation.  LB is supported by the NSF CMMT program under Grant No. DMR-2116515.
\appendix

\section{Derivation of operator product expansions}
\label{sec:OPE}

In order constrain terms in the operator product expansion \eqref{eq:OPE-def}, we first consider the allowed $\Uone_\mathrm{top}$ quantum numbers (i.e. the topological charges). In particular, the operator product $\Phi^\dagger_a \Phi_b$ has 0 net topological charge so that it is expanded in $\Uone_\mathrm{top}$ singlets. Next, we consider the $\SOsix$ tensorial structure.
A first step to decomposing the product of two $\SOsix$ tensor representations is to consider corresponding Young Tableaux.
However, we note that Young diagrams for the orthogonal groups are in general not irreducible, as (partial) traces can be subtracted.
\subsection{Monopole OPE}
It is easily seen that the monopole-monopole operator product $\Phi_a \Phi_b$ possesses a topological charge of 2, such that its expansion may only contain higher-order monopole operators which are notably less relevant and thus excluded from our study.

For the monopole-antimonopole OPE, both transforming as $\SO(6)$ vectors, we find
\begin{equation}
    \ydiagram{1} \otimes \ydiagram{1} = \ydiagram{2} \oplus \ydiagram{1,1},
\end{equation}
where $\ydiagram{2}$ further splits into the rank-2 traceless symmetric tensor (20-dimensional) and trivial irreps.
The 15-dimensional antisymmetric $\ydiagram{2}$ representation is isomorphic to the 15 $\SUfour$ adjoint masses. An explicit isomorphism is constructed by writing
\begin{equation}
    \Phi_a^\dagger \Phi_b \sim \frac{i}{2} \Phi_\alpha^\dagger L^{ab}_{\alpha \beta} \Phi_\beta + \dots,
\end{equation}
where $L^{ab}_{\alpha \beta} = - \iu (\delta^a_\alpha \delta^b_\beta- \delta^a_\beta \delta^b_\alpha)$ are $\SOsix$ generators in the defining representation.
We can identify two mutually commuting subalgebras $\{L^{23},L^{31},L^{12}\}$ and $\{L^{56},L^{64},L^{45}\}$ which generate two copies of $\SO(3)$.
Explicitly comparing the symmetry transformations of $\bvec{\Phi}^\dagger L^{ab} \bvec{\Phi}$ (recall that the bold font denotes the 6-component vector such that $L^{ab}$ acts as a matrix, with summation over $\alpha,\beta$ is implied) and the adjoint masses $M_{\mu\nu}$, we find that the first set may be identified with the $\SO(3)_\mathrm{valley}$ generators $M_{0i}$ (where $i=1,2,3$) and the latter with $M_{i0}$ which generate $\SO(3)_\mathrm{spin}$.
We further find that each set $\{L^{i,4}\}_{i=1,2,3}$, $\{L^{i,5}\}_{i=1,2,3}$, $\{L^{i,6}\}_{i=1,2,3}$ furnish a vector representation of $\SO(3)_\mathrm{valley}$ and $\{L^{1,j}\}_{j=4,5,6}$ etc. transform as vectors under $\SO(3)_\mathrm{spin}$, such that the mixed generators $\bvec{\Phi}^\dagger L^{ij} \bvec{\Phi}$ transform as $M_{j-3,i}$ where $i=1,2,3$ and $j=4,5,6$.
Note that the $C_6$ and $\mathcal{T}$ symmetries which reverse the $\Uone_\mathrm{top}$ charges of monopole operators further imply that this identification holds up to a \emph{real} constant, implying $c_{\Phi\Phi}^M \in \mathbb{R}$ in \eqref{eq:phi-phi-OPE}.

This mapping of $\SO(6)$ fundamental generators to $\SUfour$ adjoints can be conveniently expressed in terms of a tensor $\mathcal{F}$ (and its inverse $\bar{F}$ defined via $\mathcal{F}^{ab}_{\mu \nu} \bar{\mathcal{F}}^{\rho \lambda}_{ab} = \delta^{\rho}_\mu \delta^{\lambda}_\nu$) which allows us to write $L^{ab} = \mathcal{F}^{ab}_{\mu \nu} M^{\mu \nu}$ and $M^{\mu \nu} = \bar{\mathcal{F}}^{\mu \nu}_{ab} L^{ab}$ (note that all sums over $\SUfour$ adjoint indices $(\mu,\nu)$ are taken to imply $\mu = \nu = 0$ is excluded).
Matrix elements of $\mathcal{F}$ can be read off explicitly given the above mapping, and we find that the non-zero components are of the form
\begin{align} \label{eq:Fsymbol_def}
    \mathcal{F}^{ab}_{0i} = \epsilon^{abj} \ &\text{for} \ a,b \leq 3, \quad  \mathcal{F}^{ab}_{i0} = \epsilon^{a-3,b-3,i} \ \text{for} \ a,b \geq 4 \nonumber\\
    &\text{and}\quad\mathcal{F}^{ab}_{ij} = \delta^{a}_j \delta^{b}_{i+3} - \delta^a_{i+3} \delta^b_j,
 \end{align}
where the latin indices take values $i,j = 1,2,3$.
An explicit calculation shows that one may take $\bar{\mathcal{F}}^{\mu \nu}_{ab} = \mathcal{F}_{\mu \nu}^{ab} /2$.

\subsection{Monopole-mass OPE}

Next, we consider the monopole-mass operator product $\Phi_a M_{\mu \nu}$ which has unit topological charge, suggesting that unit-flux monopole operators appear in the expansion. In order to decompose the product under $\SOsix \simeq \SUfour$, it is convenient to use the previously established mapping $M^{\mu \nu} = \bar{\mathcal{F}}^{\mu \nu}_{ab} L^{ab}$ and decompose the product $\Phi_a L^{bc}$ instead, with Young tableaux
\begin{equation}
    \ydiagram{1} \otimes \ydiagram{1,1} = \ydiagram{1,1,1} \oplus \ydiagram{2,1},
\end{equation}
where the 70-dim. $\ydiagram{2,1}$ is further decomposed by subtracting a trace over the horizontal boxes in the second diagram, such that $70 \to 64 + 6$.
Note that this result is corroborated by the $\SUfour$ Young tableaux (which are irreducible)
\begin{equation}
    \ydiagram{2,1,1} \otimes \ydiagram{1,1} = \ydiagram{3,2,1} \oplus \ydiagram{2} \oplus \ydiagram{2,2,2} \oplus \ydiagram{1,1},
\end{equation}
with the last diagram on the right-hand side corresponding to the $6$-dimensional rank-2 antisymmetric representation of $\SUfour$.
Explicitly, one can decompose
\begin{align}
	&\Phi_a L^{bc} = \frac{1}{3} \left(\Phi_{a} L^{bc} + \mathrm{cyclic}\right)  \nonumber\\
	&+ \frac{1}{3} \left( 2 \Phi^a L^{bc} + \Phi_b L^{ac} - \Phi_c L^{ab} -\frac{3}{5} T^{abc} \right) \nonumber\\
	&+\frac{1}{5} \underbrace{\left(\delta^{ac} \Phi_m L^{bm} -\delta^{ab}  \Phi_k L^{ck} \right)}_{=:T^{abc}} \label{eq:decomp-Phi-L}
\end{align}
Using the explicit transformations in table \ref{tab:sym}, we find that $\Phi_m L^{bm}$ transforms as $\sim \iu \Phi_b$ with a \emph{real} constant of proportionality (note that the complex phase is constrained by $C_6$ and $\mathcal{T}$-symmetries).
The additional Kronecker-deltas in \eqref{eq:decomp-Phi-L} keep the antisymmetry in $b,c$ manifest.
Using above introduced $\mathcal{F}$-symbols, we thus find $\Phi^a M^{\mu \nu} = \bar{\mathcal{F}}^{\mu \nu}_{bc} \Phi^a L^{bc} \sim \bar{\mathcal{F}}^{\mu \nu}_{ba} \iu \Phi^b - \bar{\mathcal{F}}^{\mu \nu}_{ac} \iu \Phi^c$.
Using the antisymmetry of $\bar{\mathcal{F}}^{\mu \nu}_{ac} = -\bar{\mathcal{F}}^{\mu \nu}_{ca}$, Eq.~\eqref{eq:phi-M-OPE} with some $c^{\Phi}_{\Phi M} \in \mathbb{R}$ follows.

\section{Quartic terms in perturbation}
\label{app:spin}

In this appendix, we extend the line of thought sketched in \ref{sec:perturbativeMFT} and determine how the perturbative effective action at quartic order lifts the degeneracy of linearly combining $h_1^a(0)$ and $h_2^a(0)$ at the mean-field saddlepoint (found at quadratic order).
Note that while the effective free energy $F_\mathrm{mf}[\bvec{h}_1,\bvec{h}_2] = F_\mathrm{mf}[\bvec{h}_1] + F_\mathrm{mf}[\bvec{h}_2]$ is linear in the layer index, $F_\mathrm{mf}[\bvec{h}_1]$ depends on $\bvec{h}_2$ (or vice versa) once we plug in \eqref{eq:A} which holds as a solution of the self-consistency equations at the mean-field saddlepoint. 

The free energy up to quartic order is
\be
F\mf[h] = - \log \mathcal{Z}_0 - \frac{\langle \mathcal{S}_h^2 \rangle_0}{2}+ \frac{\langle \mathcal{S}_h^2 \rangle_0^2}{8} - \frac{ \langle \mathcal{S}_h^4 \rangle_0}{24} + \dots,
\ee
where as usual, $\langle \cdot \rangle_0$ denotes evaluating expectation values in the unperturbed compact \qed{}.
The first two terms on the right hand side have no preference on the relationship between $h_1^a(0)$ and $h_2^a(0)$ because of the linearity of the self-consistency equations at weak coupling.

The expression for $\bk{\mathcal{S}_h^2}_0$ is relatively straightforward, with 
\be
\bk{\mathcal{S}_h^2}_0=2\sum_{l,a}\int \du^3x \du^3y\  \frac{h_l^a(x)h_l^{a*}(y)}{|x-y|^{2\Delta_\Phi}},
\label{eq:Sh2}
\ee
where we have omitted the OPE coefficients for convenience.  
Plugging in the expressions that we found, 
\be
\begin{split}
& a\leq 3: \begin{cases}
h_1^a(\x)= \tilde{h}_1^a(0)-A\tilde{h}_2^a(0)\fs(\x),\\
 h_2^a(\x)= \tilde{h}_2^a(0)-A\tilde{h}_1^a(0)\fs^{\ast}(\x), 
\end{cases}\\
& a\geq 4: \begin{cases}
h_1^a(\x)= \tilde{h}_1^a(0)-A\tilde{h}_2^a(0)\fv^a(\x),\\
 h_2^a(\x)= \tilde{h}_2^a(0)-A\tilde{h}_1^a(0)\fv^{a\ast}(\x), 
\end{cases}
\end{split}
\label{eq:h_ansatz}
\ee
and upon evaluating the integrals, one easily observes that the leading terms in \eqref{eq:Sh2} scale as
\be
\bk{\mathcal{S}_h^2}_0\sim  L^{6-2\Delta_\Phi} (|\tilde{\bm{h}}_1(0)|^2+|\tilde{\bm{h}}_2(0)|^2),
\label{eq:Sh2_leading}
\ee
where we define the norm of the (complex) $\SO(6)$ vectors $|\tilde{\bm{h}}_l(0)|^2=\sum_{a} \tilde{{h}}_l^a(0)\tilde{{h}}_l^{a*}(0)$.
From \eqref{eq:Sh2_leading}, it is clear that the $\bk{\mathcal{S}_h^2}_0$ and $\bk{\mathcal{S}_h^2}_0^2$ (and any higher powers of $\langle \mathcal{S}_h^2\rangle_0$) together pin down the total magnitude $(|\tilde{\bm{h}}_1(0)|^2+|\tilde{\bm{h}}_2(0)|^2)$, but will not determine the relative ratio $|\tilde{\bm{h}}_1(0)|^2/|\tilde{\bm{h}}_2(0)|^2$. 

Hence, we now turn to the contribution $\langle \mathcal{S}_h^4 \rangle_0$.
Using the OPEs \eqref{eq:phi-phi-OPE}\eqref{eq:phi-M-OPE}, the leading terms come from the fusion channels 
$(\Phi^\dagger \times \Phi)\times (\Phi^\dagger \times \Phi)\rightarrow M\times M\rightarrow 1$:
\be
\begin{split}
\langle \mathcal{S}_h^4 \rangle_0=& -\sum_l \sum_{a\neq b}\int \du^3 x\ \du^3y\ \du^3z\ \du^3 w\   \\
&  \frac{1}{|x-y|^{2\Delta_\Phi-\Delta_M}}\frac{1}{|y-z|^{2\Delta_M}} \frac{1}{|z-w|^{2\Delta_\Phi-\Delta_M}}\\
& \times [h_l^{a\ast}(x)h_l^{b}(y)h_l^{a\ast}(z)h_l^b(w) \\
& \quad \quad -h_l^{a\ast}(x)h_l^{b}(y)h_l^{b\ast}(z)h_l^a(w)]+\hc.
\end{split}
\ee
The expression above contains various combinations of quartic terms $\tilde{h}_l^*(Q_1)\tilde{h}_l(Q_2)\tilde{h}_l^*(Q_3)\tilde{h}_l(Q_4)$, with $Q_i$ being either zero or some finite momentum wavevector appearing in $\fs(x)$ or $\fv^a(x)$. Naively one would expect the contributions from the uniform piece $\tilde{h}_l^*(0)\tilde{h}_l(0)\tilde{h}_l^*(0)\tilde{h}_l(0)$ to dominate, as the corresponding integrals contain the biggest power of the IR cutoff $L$, and scales as $J_\alpha^4 L^{12-4\Delta_\Phi}$.
However, the factor $A$ which appears in \eqref{eq:h_ansatz} defined in \eqref{eq:A} contributes non-trivial scaling, $A \sim (L/a_\mathrm{m})^{3/2-\Delta_\Phi}$.
Carefully taking this into consideration, we find the dominating terms to be ones of the form $\tilde{h}_l^*(0)\tilde{h}_l(Q_1)\tilde{h}_l^*(Q_2)\tilde{h}_l(0)$ with $Q_1, Q_2\neq 0$.
They scale like $ A^2 L^{6-4\Delta_\Phi +2\Delta_M} |q|^{2\Delta_M-3}\sim  L^{12-4\Delta_\Phi} (L/a_M)^{2\Delta_M-\Delta_\Phi},$ which is more divergent than $ L^{12-4\Delta_\Phi}$ as $2\Delta_M>\Delta_\Phi$, such that we expect this term to dominate in the thermodynamic limit $L\to \infty$. Combining with \eqref{eq:h_ansatz}, the leading terms can thus be derived
\be
\begin{split}
- \langle \mathcal{S}_h^4 \rangle_0 \sim &    L^{12-4\Delta} \left(\frac{L}{a_M}\right)^{2\Delta_M-2\Delta_\Phi}\\
& \times [ |\bm{h}_1(0)\cdot \bm{h}_2(0)|^2-|\bm{h}_1(0)|^2 \cdot |\bm{h}_2(0)|^2  ]
\end{split} \label{eq:quartic_h1h2}
\ee
The first term can be minimized by taking $\tilde{\bm{h}}_1(0) = s \tilde{\bm{h}}_2(0)$, with $s \in \mathbb{C}$ being a complex number.
In order to fix the relative magnitudes, it is convenient to employ a more symmetric parametrization that keeps the total magnitude fixed, $\tilde{\bm{h}}_1(0) = \bm{h} \cos\alpha$ and $\tilde{\bm{h}}_2(0) = r \bm{h} \sin\alpha$,
where $r\in \mathbb{C}$ is a complex phase, $|r|^2 = 1$, and $\bm{h}$ is some complex vector whose magnitude is renormalized by higher-order terms.
Above parametrization in terms of $\alpha$ hence keeps the $\Uone$ degeneracy in $(|\tilde{\bm{h}}_1(0)|^2+|\tilde{\bm{h}}_2(0)|^2)=\mathrm{const.}$ manifest.
Maximizing the second term \eqref{eq:quartic_h1h2} amounts to taking $\cos^2\alpha \sin^2 \alpha = 1/4$, implying that $|\tilde{\bm{h}}_1(0)|^2=|\tilde{\bm{h}}_2(0)|^2$.
Consequently, we find $\tilde{\bm{h}}_1(0) = r \tilde{\bm{h}}_2(0)$ where $r$ is a complex phase.

\section{Perturbative conformal renormalization group}\label{app:conformal-rg}

In this appendix, we briefly review the conformal renormalization group \cite{cardy96} and describe its application to the homogenous bilayer system, which serves as a preparation for the discussions in section \ref{sec:conformal-rg}.

The procedure, formulated in real space, consists in (1) raising the UV cutoff $a\to ba $ which sets a minimal separation $a\leq |x-y|$ of two operator insertions at points $x$, $y$, where $b>1$, and (2) a subsequent scale transformation which restores the action, but with modified couplings $g'$.
Explicitly, we may consider the partition function $\mathcal{Z} = \int \mathcal{D}[\{\mathcal{O}\}] \, \eu^{-\mathcal{S}_0-\mathcal{S}_g}$, where $\mathcal{S}_0$ is the action at the conformal fixed point,  $\mathcal{S}_g = \sum_i a^{\Delta_i-3} \int \du^3 x \,  g_i \mathcal{O}_i(x)$ is the perturbing action with some operator $\mathcal{O}_i(x)$ with scaling dimension $\Delta_i$. Expanding perturbatively in $g$, one may write $\mathcal{Z} = \mathcal{Z}_0 \langle 1 - \mathcal{S}_g + \mathcal{S}_g^2 + \dots \rangle_0$, with the expectation value to be taken with respect to the fixed point action and over configurations obeying the UV cutoff $a$.

We now consider the effective action $\mathcal{S}_{g}'$ obtained by introducing a new cutoff $b a$ with $b>1$, integrating over those configurations with operator insertion separations below the new cutoff, and rescaling coordinates (and operators) $x = b x'$ such that \ufps{the} effective action has again cutoff $a$.
As the linear term $\langle \mathcal{S}_g \rangle_0$ only features single operator insertions, it is unaffected from raising the cutoff and contributes the ``bare'' scale transformation of the coupling at the fixed point,
\begin{equation}
    a^{\Delta_i-3} \int \du^3 x \, g_i \mathcal{O}_i(x) = a^{\Delta_i-3} \int \du^3 x' \, g_i b^{3-\Delta_i} \mathcal{O}_i(x').
\end{equation}
At quadratic order, one has
\begin{align} \label{eq:conf_rg_quadratic}
    \langle \mathcal{S}_g \rangle_0 = &\sum_{ij} a^{\Delta_i + \Delta_j - 6} g_i g_j \Bigg[ \int_{|x-y|> ba} \du^3 x \,\du^3 y \, \langle \mathcal{O}_i(x) \mathcal{O}_j(y) \rangle_0 \nonumber\\ &+ \int_{ b a > |x-y|>a} \du^3 x \, \du^3 y \, \langle \mathcal{O}_i(x) \mathcal{O}_j(y) \rangle_0  \Bigg].
\end{align}
Upon rescaling, the first term  in \eqref{eq:conf_rg_quadratic} reproduces the ``bare'' scaling behavior of the quadratic term.
For the second term however, $\mathcal{O}_i(x)$ and $\mathcal{O}_j(y)$ are close (in the sense that their separation lies below the raised cutoff) and thus their product is replaced using their OPE \eqref{eq:OPE-def}, which contains (at leading order) a primary operator $\mathcal{O}_k$, and thus contributes to the scaling of the linear term in the expansion of $\eu^{-\mathcal{S}_g'}$. Considering $b = 1+\delta l$ with infinitesimal $\delta l$, the occurring integral over an infinitesimal shell is evaluated as $\int_{a<|x-y|<(1+\delta l)a} \du^3 x \,  |x-y|^{-\Delta_i - \Delta_j + \Delta_k} = 4 \pi  a^{3+\Delta_k- \Delta_i - \Delta_j} \delta l$ and one arrives at the differential RG equations
\begin{equation}
    \frac{\du g_k}{\du l} = (3-\Delta_k) g_k - 2 \pi \sum_{i,j} g_i g_j C_{ij}^k,
\end{equation}
where $C_{ij}^k$ are the OPE coefficients as defined in \eqref{eq:OPE-def}.
Turning to the homogenous bilayer system with Lagrangian \eqref{eq:L_aa_full}, one proceeds analogously. In practice, we find it more convenient to explicitly expand $\exp[-\int \du^3 x \mathcal{L}_\mathrm{int}^{(12)}]$ to quadratic order and then read off the respective contributions to the renormalized couplings.
Here, we note the identities
\begin{align}
    &\sum_{a,i=1}^{3} \mathcal{F}^{ab}_{0i} \mathcal{F}^{ac}_{0i} = 2 \delta^{bc} \ \text{for} \ b,c \leq 3 \\
    &\sum_{a=4}^6\sum_{i=1}^{3} \mathcal{F}^{ab}_{i0} \mathcal{F}^{ac}_{i0} = 2 \delta^{bc}  \ \text{for} \ b,c \geq 4 \\
    &\sum_{a=1}^3 \sum_{i,j=1}^3 \mathcal{F}^{ab}_{ij} \mathcal{F}^{ac}_{ij} = 3\delta^{bc} \ \text{for} \ b,c \geq 4 \\
    &\sum_{a=4}^6 \sum_{i,j=1}^3 \mathcal{F}^{ab}_{ij} \mathcal{F}^{ac}_{ij} = 3\delta^{bc} \ \text{for} \ b,c \leq 3\\
    &\sum_{a,b=1}^3 \mathcal{F}^{ab}_{\mu \nu} \mathcal{F}^{ab}_{\rho \lambda} = 2 \delta_{\mu,0} \delta_{\rho,0} \delta_{\nu,i}  \delta_{\lambda,j} \delta_{i,j}  \\
    &\sum_{a=1}^{3}\sum_{b=4}^6 \mathcal{F}^{ab}_{\mu \nu} \mathcal{F}^{ba}_{\rho \lambda} = \sum_{a=4}^{6}\sum_{b=1}^3 \mathcal{F}^{ab}_{i j} \mathcal{F}^{ba}_{m n} = -2 \delta_{i m} \delta_{j n},
\end{align}
where latin indices $i,j,\dots \in \{1,2,3 \}$ as usual, which follow straightforwardly from the explicit form of the $\mathcal{F}$-symbol given in \eqref{eq:Fsymbol_def}.
With these results, the flow equations given in \eqref{eq:flow-eqns} follow.



\bibliography{twidsl}
\bibliographystyle{apsrev4-2}

\end{document}